\documentclass[useAMS,usenatbib,letter]{mn2e}
\usepackage{graphicx}
\usepackage{amsmath}
\usepackage{amssymb}
\usepackage{amsfonts}
\usepackage{array}
\usepackage{natbib}
\usepackage[utf8]{inputenc}
\usepackage[spanish, activeacute]{babel}
\usepackage{epsfig}
\usepackage{dsfont} 
\usepackage{fancyhdr}
\usepackage{ucs}
\usepackage{color}
\usepackage{longtable}
\usepackage{calc}
\usepackage{multirow}
\usepackage{hhline}
\usepackage{ifthen}
\usepackage{lscape}
\def\et{et al. }

\title[Regions of Dynamical Stability for Discs and Planets in Binary
  Stars of the Solar Neighborhood]{Regions of Dynamical Stability for
  Discs and Planets in Binary Stars of the Solar Neighborhood}

\author[L. G. Jaime B. Pichardo L. Aguilar]{Luisa G. Jaime$^{1,2}$,
  Barbara Pichardo$^{1}$ \thanks{E-mail: barbara@astroscu.unam.mx
    (BP)}, and Luis Aguilar$^{3}$\\ $^{1}$Instituto de Astronom\'\i a,
  Universidad Nacional Aut\'onoma de M\'exico, Apdo. postal 70-264,
  Ciudad Universitaria, M\'exico\\ $^{2}$Instituto de Ciencias
  Nucleares, Universidad Nacional Aut\'onoma de M\'exico, Apdo. postal
  70-543 Ciudad Universitaria, D.F., M\'exico \\ $^{3}$Instituto de
  Astronom\'ia, Universidad Nacional Aut\'onoma de M\'exico,
  Apdo. postal 877, 22800 Ensenada, M\'exico}

\begin{document}

\date{Accepted xxx. Received ; in original form }

\pagerange{\pageref{firstpage}--\pageref{lastpage}} \pubyear{}

\maketitle

\label{firstpage}

\begin{abstract} 
Using the results of Pichardo \et (2005,2008), we determine regions of
dynamical stability where planets (or discs in general) could survive
in stable orbits around binary stellar systems. We produce this study
for 161 binary stars in the Solar neighborhood with known orbital
parameters. Additionally, we constructed numerically the discs
(invariant loops) around five binary systems with known orbital
parameters and with confirmed planets: HIP 10138, HIP 4954, HIP 67275,
HIP 116727 and Kepler 16, as a test to the approximation of Pichardo
et al. (2005,2008). In each single case, the reported position of the
planets lay within our calculated stability regions. This study
intends to provide a guide in the search for planets around binary
systems with well know orbital parameters, since our method defines
precise limits for the stable regions, where discs may have
established and planets formed.

\end{abstract}

\begin{keywords}
circumstellar matter, discs -- binary: stars, Solar Neighborhood, exoplanets.
\end{keywords}

\section{Introduction}\label{Intro}
It is known that most low-mass main-sequence stars are members of
binary or multiple systems (Duquennoy \& Mayor 1991; Fisher \& Marcy
1992), and in particular in the Solar Neighborhood, the fraction goes
up to $\sim 78\%$ (Abt 1983). This suggests that binary formation is
the primary branch of the star formation process (Mathieu 1994).

Significant advances in high-angular-resolution infrared imaging
technology have enabled large surveys of young binary stars on a
variety of star-forming regions (Mathieu \et 1992, 1994). In addition,
right after the discovery of the first extrasolar planetary system
around a pulsar (Wolszczan \& Frail 1992), and particularly after the
first extrasolar planet discovered around a main sequence star (Mayor
\& Queloz 1995; Marcy \& Butler 1998), observational activity was
greatly stimulated. More recently, advances in observational
techniques and instrumentation, such as the HST (WFPC2 \& NICMOS)
imaging (Padgett \et 1997, 1999; Reid \et 2001; Borucki \et 2010),
submillimeter imaging (Smith \et 2000), optical and infrared
long-baseline interferometry (Quirrenback 2001a,b), millimeter and
submillimeter interferometry (Launhardt \et 2000, Launhardt 2001,
Guilloteau 2001), adaptive optics (Simon \et 1999; Close 2001),
spatial astrometry (S\"oderhjelm 1999; Quist \& Lindegren 2000, 2001),
and microlensing (Alcock \et 2001; Dong-Wook \et 2008; Rattenbury
2009), are also available in binary studies. Thanks to all this new
technology and observational work, we have now the possibility to
study and understand better the physics of binary systems and the
surrounding discs built during the formation stage.

In the recent past, several planets in binary or multiple star systems
have been discovered (Correia 2008; Deeg \et 2008; Desidera \&
Barbieri 2007; Fischer \et 2008; Raghavan 2006; Konacki 2005; Mugrauer
\et 2005; Eggenberger \et 2004; Sigurdsson \et 2003; Udry \et 2002;
Sigurdsson \& Phinney 1993; Lyne 1988, etc.). Both suspected of formed
in {\it situ}, or acquired by dynamical processes (Pfahl \&
Muterspaugh 2006). For a review of observational techniques see
Muterspaugh \et (2010). In addition, several recently discovered
circumstellar discs where planets are assumed to be formed, lie around
close binaries (Wright \et 2011; Prato \& Weinberger 2010; Desidera \&
Barbieri 2007; Doyle \et 2011; Queloz \et 2000; Hatzes \et 2003). In
these cases the presence of a companion star should have a very strong
influence on both discs and planets. From the several hundreds of
extrasolar planets confirmed so far (see http://exoplanet.eu,
http://planetquest.jpl.nasa.gov, www.exoplanets.org) to be around main
sequence stars, about 10\% are known to reside in binary systems with
a wide range of orbital separations. In almost all cases, the planet
orbits in S-type configurations (Dvorak 1986), while the second star
acts as a perturber to the planetary system. A circumbinary planet
(P-type orbit) has recently also been detected in Kepler 16 (Doyle et
al. 2011). This has motivated the search for stable periodic orbits
around binary systems where planets (and discs in general) can settle
down in a stable configuration. Most theoretical studies have focused
on binaries in near-circular orbits (H\'enon 1970; Lubow \& Shu 1975;
Paczy\'nski 1977; Papaloizou \& Pringle 1977; Rudak \& Paczy\'nski
1981; Bonell \& Bastien 1992; Bate 1997; Bate \& Bonnell 1997). Even
very precise analytical methods to approximate periodic orbits in
circular binaries are available (Nagel \& Pichardo 2008).

Due to the lack of conservation of the Jacobi integral, the case of
eccentric binaries is qualitatively more complicated. Artymowicz \&
Lubow (1994) and Pichardo \et (2005, 2008, hereafter PSA1 and PSA2)
calculate the extent of zones in phase space available for stable, non
self-intersecting orbits around each star and around the whole
system. In this study we use the results of PSA1 and PSA2, to
calculate stable regions for planets or discs in binary systems in the
Solar Neighborhood with known orbital parameters such as, mass ratio,
eccentricity and semimajor axis.

Although the existence of stable zones, as the ones we are calculating
here, is a necessary, but not a sufficient condition to the existence
of planets (or discs in general), if any stable material (planets,
gas, etc.), exists in a given system, irrespective of their formation
mechanism, they would be necessarily located within the limits of the
stable zones. In this direction, a fruitful line of investigation is
the intersection between phase space available zones for the long term
evolution of planetary systems and habitable regions allowed by the
binary system (Haghighipour \et 2007; Haghighipour \et 2010).

We present a table with the compilation of all the binaries in the
Solar Neighborhood with known orbital parameters from different
sources. In the same table are presented the results for our
calculated circumstellar and circumbinary stable zones around each
binary system.

This paper is organized as follows. In Section \ref{method}, we
explain briefly the method employed to calculate regions of stable non
self-intersecting orbits around binary stars. The binary star sample
is presented in Section \ref{sample}. Results, including stable
regions of orbital stability for circumstellar and circumbinary
planetary discs (and discs in general), and an application to
observations of five binaries with observed planets, are given in
Section \ref{results}. In Section \ref{conclusions} we present our
conclusions.

\section{The Method}\label{method} 

In the simpler case of circular binary orbits, the potential is
time-independent in the co-rotating frame and thus the Jacobi energy
is conserved. This allows the existence of closed orbits that, when
stable, spawn the orbital structure of the system (Carpintero \&
Aguilar 1998). In the general case of binaries in eccentric orbits,
the problem is more complex, as now the potential is
time-dependent. However, we can exploit the fact that the
time-dependency is strictly periodic.

A time-periodic potential in 2-D can be casted as a 3-D system with an
autonomous hamiltonian, with the addition of time and the original
hamiltonian as two extra dimensions in phase space. Regular orbits
will lie on 3-D manifolds and be multiple periodic with three
frequencies, one of which is given by the binary orbital frequency. If
we take snapshots at a fixed binary phase, the projections of a regular
orbit will lie on a 2-D manifold. If the orbit has an additional
isolating integral of motion, this projection will now lie on a 1-D
manifold: an "invariant loop" (Maciejewski \& Sparke 1997, 2000).

\begin{figure}
\includegraphics[width=74mm]{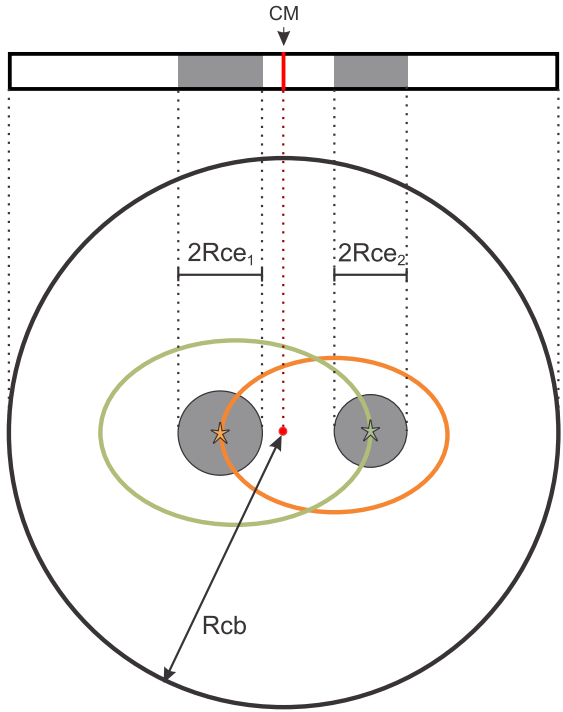}
\vspace{0.1cm}
\caption{Schematic figure of the geometry of the stable zones in a
  binary system. The black outer circle represents the inner boundary
  of the circumbinary region (eq. \ref{eqCB}). The orange and green
  curves are the primary and secondary stars. The gray circular areas
  around each star represent the circumstellar stable zones
  (eq. \ref{eqCE}). The red dot at the center represents the system
  barycenter. In the upper part we show a schematic rectangle that we
  will use in Table \ref{tabla_muestra}, to show graphically the
  relative extent and situation of the stable regions, in relation to
  the stars, for each entry in our sample. Dotted lines join the
  corresponding parts in the upper and lower representations.}
\label{fig.binscheme}
\end{figure}

Stable invariant loops represent the generalization to periodically
time-varying potentials of stable periodic orbits in steady
potentials. PSA1 and PSA2, implemented a numerical method to find
them. The equations of motion are solved in an inertial reference
frame with Cartesian coordinates with the origin at the binary
barycenter. An ensemble of test particles is launched when the binary
star is at periastron, and from the line that joins both stars at that
moment, to search for invariant loops. A more detailed explanation of
the method and a study of the phase space in binary systems is in
those references. In this work we employ the formulae from PSA1
(Eq. 6) and PSA2 (Eq. 6). These relations provide the maximum radius
of circumstellar stable zones and the inner radius of the circumbinary
stable zone, in terms of the mass ratio ($q = m_2 / (m_1 + m_2)$,
where $m_1$ and $m_2$ are the masses of the primary and secondary
stars, respectively), and the eccentricity ($e = \sqrt{1-b^2/a^2}$,
where $a$ and $b$ are the semimajor and semiminor axes of the binary
orbit). The radius of the outer limit of the circumstellar stable
zones from PSA1, is given by,

\begin{eqnarray}
R_i =  R_{i,Egg}\ \times \ [0.733 (1-e)^{1.20} q^{0.07}] \, ,
\label{eq2.1}
\end {eqnarray}
and a similar study in PSA2 but for the circumbinary region, gives a relation
for the inner radius,

\begin{eqnarray}
R_{CB}(e,q) \approx 1.93 \ a \ (1 + 1.01e^{0.32})\ [q(1-q)]^{0.043}.
\label{eqCB}
\end {eqnarray}
In these relations $R_{i,Egg}$ is the approximation of Eggleton to the
maximum radius of a circle circumscribed within the Roche lobe
(Eggleton 1983):

\begin{eqnarray}
R_{i,Egg}/a = \frac{0.49 q_i^{2/3}}{0.6q_i^{2/3}+ln(1+q_i^{1/3})},
\label{eqCE}
\end {eqnarray}

and

\begin{eqnarray}
q_1 = m_1/m_2 = \frac{1-q}{q} \ \ {\rm and \ \ } q_2 = m_2/m_1 = \frac{q}{1-q}.
\label{qEgg}
\end {eqnarray}
In Figure \ref{fig.binscheme} we present a schematic figure of the
geometry of the system. We show in this figure some of the variables
involved in the problem.

We must emphasize that the regions of stable orbits we report here,
are the regions were these invariant loops exist, and furthermore, we
restrict ourselves only to non self-intersecting orbits, where gas
could settle and planets may form. It is in this sense that the term
"stable region" should be understood in this work. Using these
formulae and restriction, we have calculated circumbinary and
circumstellar radii for our sample with a total of 161 binary systems
in the solar neighborhood with known orbital parameters.

\section {The sample}\label{sample}
The previous equations require, besides the stellar mass ratio, the
semimajor axis and orbital eccentricity for each binary system, two
parameters that are difficult to determine observationally. There is a
diversity of methods that have been used to estimate them. Our sample
is a compilation from different sources (Jancart \et 2005; Martin \et
1998; Strigachev \& Lampens 2004; Bonavita \& Desidera 2007; Holman \&
Wiegert 1999; Mason \et 1999; Latham \et 2002; Balera \et 2006;
Cakirli \et 2009; Milone \et 2005; D\'\i az \et 2007; Konacki \et
2010; Desidera \& Barbieri 2007; Doyle \et 2011). We include all
binaries with an estimation of these parameters within 100 pc from the
Sun (currently 161 systems).

\begin{landscape}
\begin{figure*}
\begin{tabular}{llllllllllc}
\hline
Object		&Alter. name&	 $a$	&      $e$	&    	$M_{1}$		&  	$M_{2}$		&	$R{ce1}$	&	$R{ce2}$	&	$R_{cb}$	&Ref&	Scheme	\\
 HIP		&	&       (AU)	&		&	(M$_{\odot}$)		&	(M$_{\odot}$)		&	 (AU)		&  	(AU)		&  	(AU)		& &  	 $*_{1}$  \   $*_{2}$ \\
\hline
\bf{116310}	&	\bf{HD 221673}	&	\bf{95}	&	\bf{0.322}	&	\bf{2.0}	&	\bf{2.0}	&	\bf{15.70}	&	\bf{15.70}	&	\bf{294.14}	&	\bf{16}	&		\includegraphics[width=45mm]{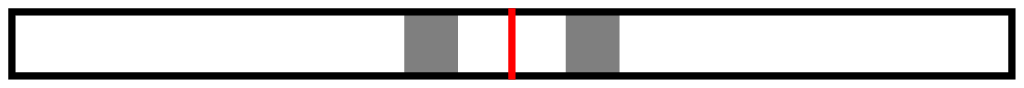} \\

\bf{116727}	&	\bf{HD 222404}	&	\bf{18.5}	&	\bf{0.36}	&	\bf{1.59}	&	\bf{0.4}	&	\bf{3.55}	&	\bf{1.9}	&	\bf{57.04}	&	\bf{13}	&		\includegraphics[width=45mm]{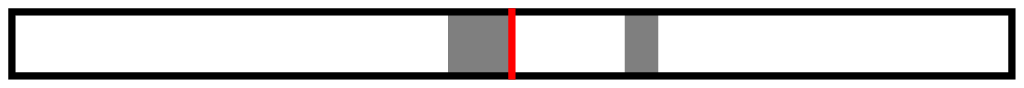} \\

117666	&	WDS 23517+0637	&			10.2	 		&		0.3	 		&			0.6	 		&			0.58	 		&		1.77	 	&		1.74	 	&		31.29	 	&	6	&		\includegraphics[width=45mm]{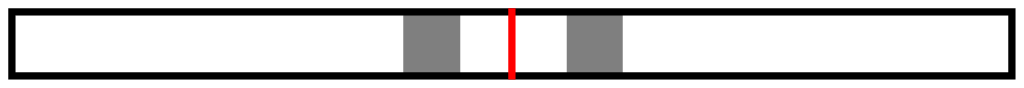} \\

\bf{}	&	\bf{Kepler 16}	&	\bf{0.22}	&	\bf{0.16}	&	\bf{0.69}	&	\bf{0.20}	&	\bf{0.06}	&	\bf{0.03}	&	\bf{0.63}	&	\bf{14}	&		\includegraphics[width=45mm]{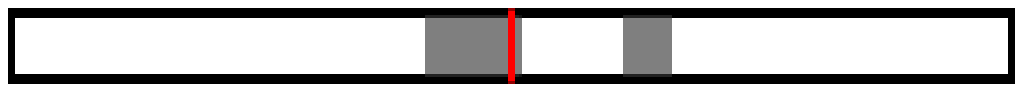} \\

         &       K10848064          &      0.049     &   0      &     1.2     &     0.073   &    0.018   &   0.005   &    0.083    &    15      &   \includegraphics[width=45mm]{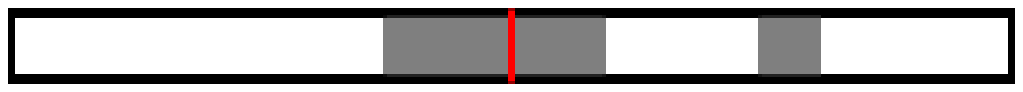}\\
\\
         
         &       K08016222          &      0.065     &   0.044  &     1.1     &     0.086   &    0.023   &   0.007   &    0.154    &    15      &   \includegraphics[width=45mm]{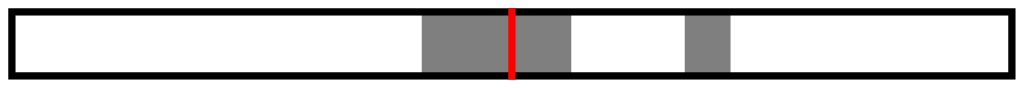}\\
          
         &       K09512641          &      0.060     &   0      &     1.2     &     0.140   &    0.021   &   0.008   &    0.105    &    15      &   \includegraphics[width=45mm]{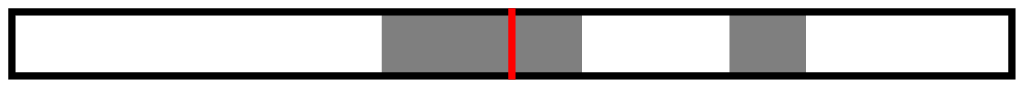}\\
         
         &       K07254760          &      0.042     &   0      &     1.2     &     0.215   &    0.014   &   0.007   &    0.075    &    15      &   \includegraphics[width=45mm]{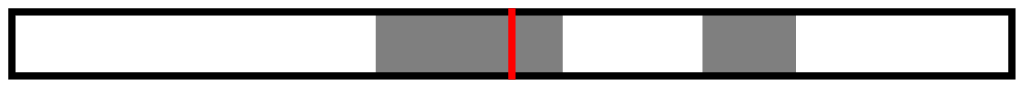}\\
         
         &       K05263749          &      0.055     &   0      &     1.3     &     0.266   &    0.018   &   0.009   &    0.097    &    15      &   \includegraphics[width=45mm]{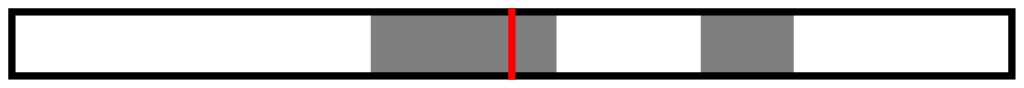}\\
        
         &       K04577324          &      0.039     &   0      &     1.2     &     0.241   &    0.013   &   0.006   &    0.069    &    15      &   \includegraphics[width=45mm]{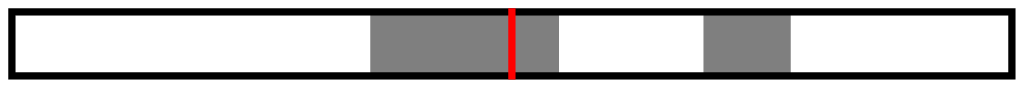}\\
\\
         
         &       K06370196          &      0.061     &   0      &     1.3     &     0.359   &    0.020   &   0.011   &    0.108    &    15      &   \includegraphics[width=45mm]{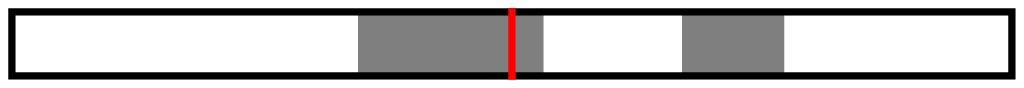}\\

\hline
\label{tabla_muestra}
\end{tabular}
(1) Jancart \et 2005, (2) Martin \et. 1998, (3) Strigachev and Lampens
2004, (4) Bonavita and Desidera 2007, (5) Holman and Wiegert 1998, (6)
Mason \et 1999, (7) Latham \et 2002, (8) Balega \et 2006, (9) Diaz \et
2007, (10) Cakirli \et 2009, (11) Milone \et 2005, (12) Konacki \et
2010, (13) Desidera and Barbieri 2007, (14) Doyle \et 2011, (15)
Faigler \et 2011, (16) Muterspaugh \et 2010
\end{figure*}
\end{landscape}

In Table \ref{tabla_muestra} we present our sample. Columns 1 and 2,
are the name of the systems in the Hipparchos catalogue and an
alternate name, if it exists. Columns 3 to 6 are our input data: the
semimajor axis, orbital eccentricity and stellar masses as reported in
the references listed in column 10. Columns 7 to 9 list our results:
the circumprimary, circumstellar, and circumbinary boundaries of the
stable regions. In the circumstellar cases, the value is the radius of
the outer boundary. In the circumbinary, is the radius of the inner
boundary. All lengths are given in AU and masses in solar masses.
Finally, the last column is a shcematic figure that gives the relative
sizes and positions of the stable zones (see Figure
\ref{fig.binscheme}). For instance, for BD -8$^{\rm o}$ 4352 (9th entry in
the table) the circumstellar regions are symmetric and cover a good
fraction of the inner hole of the circumbinary region, this is is due
to the low eccentricity of the system and its high $q$(=0.5). In
contrast, the binary called Gamma Vir (6th entry in the table),
present circumstellar regions slightly asymmetric, due to the small
mass difference between the stars, and quite narrow, due to its high
eccentricity.

\section{Results}\label{results}

The presence of a stellar companion, particularly in an eccentric
orbit, severely curtails the size and shape of the stable zones. While
a single star has circular stable orbits at all radii (neglecting
finite stellar size and tidal distortion effects), the presence of a
stellar companion severely curtails the region where stable, non-self
intersecting orbits may exist, both in extent and, to a lesser extent,
shape.

It is unclear at present the way in which these effects impose
restrictions in the process of star formation. What is clear is that
the effect is in the sense of inhibiting, rather than promote it.

From the observational side, the statistics of the observed systems
suggests that binarity does indeed has en effect on planetary masses
and orbits (Eggenberger \et 2004), even restricting terrestrial planet
formation for binary periastron smaller than 5 AU affecting discs to
within $\sim$1 AU of the primary star (Quintana \et 2007, Quintana \&
Lissauer 2010).

The goal of this study is to determine the extent of circumstellar and
circumbinary regions of stable, non-self intersecting orbits, as
plausible regions where planets could have formed and may exist. It
could also indicate possible regions of planetary formation.

\begin{figure}
\includegraphics[width=84mm]{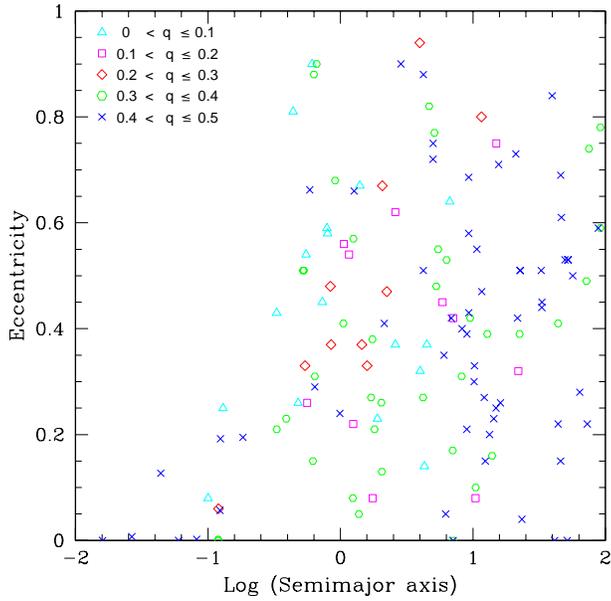}
\vspace{0.1cm}
\caption {Binary semimajor axes (AU) vs eccentricities of our
  sample. The mass ratio is indicated with various colors and symbols,
  as shown in the upper left corner.}
\label{fig.sampleSN}
\end{figure}

\begin{figure}
\includegraphics[width=84mm]{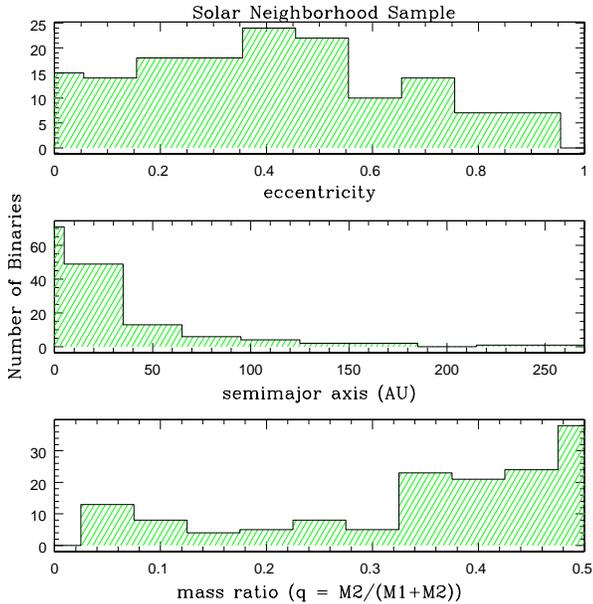}
\vspace{0.1cm}
\caption {Histograms (from top to bottom): Eccentricity, semimajor 
axis and mass ratios, for all systems in our sample.}
\label{fig.histSNOBS}
\end{figure}

\begin{figure}
\includegraphics[width=84mm]{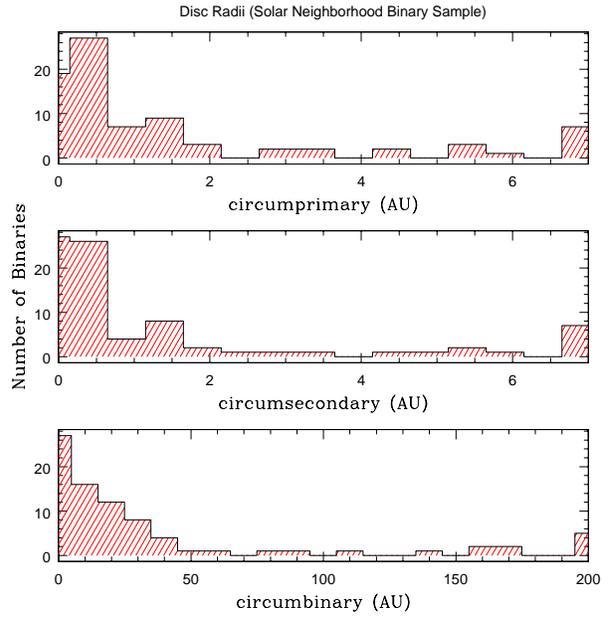}
\vspace{0.1cm}
\caption {Histograms (from top to bottom): Circumprimary, circumsecondary 
and circumbinary radii, for all systems in our sample}
\label{fig.histSNrad}
\end{figure}

Figure \ref{fig.sampleSN} shows the binary semimajor axes {\it vs.}
the orbital eccentricity of our entire sample, split in mass ratio
intervals.  As it is already known, the region of small semimajor axes
and high eccentricites is depleted, i.e. very close binaries, tidally
locked in general, have eccentricities close to zero (Duquennoy \&
Mayor 1991). A large percentage (about $60\%$) of binaries in the
sample have low eccentricities ($e\la 0.5$), small semimajor axes
($a\la 50$ AU) and large mass ratios ($q\ga 0.4$), as shown in the
histograms in Figure \ref{fig.histSNOBS}. Consequently (as seen in
Figure \ref{fig.histSNrad}), circumstellar stable regions in this
sample have small radii ($\la 2$ AU), with both stable regions
(circumprimary and circumsecondary) in most systems having similar
size. On the other hand, the majority of circumbinary gaps have radii
smaller that $\sim$50 AU.

\begin{landscape}
\begin{figure*}
\begin{tabular}{llllllllllc}
\hline
Object	&	Alter. name	&	 		$a$			&      		$e$			&    			$M_{1}$			&  			$M_{2}$			&		$R{ce1}$	&		$R{ce2}$	&		$R_{cb}$	&	Ref	&	Scheme	\\
 HIP	&			&       		(AU)			&					&			(M$_{\odot}$)		&			(M$_{\odot}$)		&		 (AU)		&  		(AU)		&  		(AU)		& 		&  	 $*_{1}$  \   $*_{2}$ \\
\hline

--	&	HD 10361   	&			52.2	 		&		0.53	 		&			0.77	 		&			0.75	 		&		5.61	 	&		5.54	 	&		173.15	 	&	4	&		\includegraphics[width=45mm]{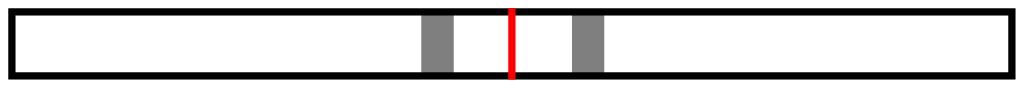} \\
--	&	HD 145958A  	&			124	 		&		0.39	 		&			0.9	 		&			0.89	 		&		18.17	 	&		18.08	 	&		393.95	 	&	4	&		\includegraphics[width=45mm]{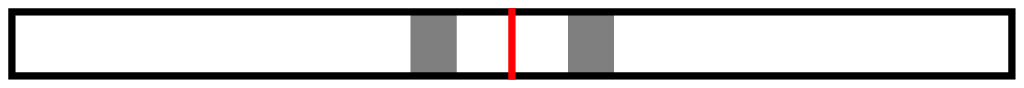} \\
--	&	HD 145958B  	&			124	 		&		0.39	 		&			0.89	 		&			0.9	 		&		18.17	 	&		18.08	 	&		393.95	 	&	4	&		\includegraphics[width=45mm]{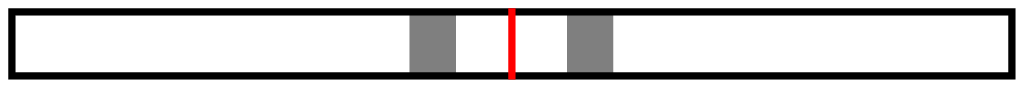} \\
--	&	HD 146362   	&			130	 		&		0.76	 		&			2.19	 		&			1.12	 		&		6.98	 	&		5.14	 	&		452.9	 	&	4	&		\includegraphics[width=45mm]{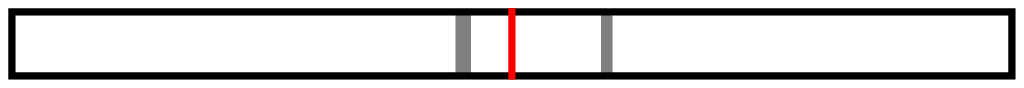} \\
--	&	$\epsilon$ Cet 	&	1.57 		&		0.27	&		1.3	 		&			1.3	 		&		   	0.28			&		0.28	 	&		4.75	 	&	5			&				\includegraphics[width=45mm]{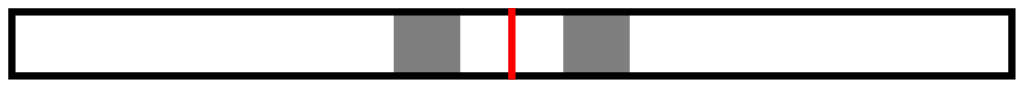} \\
\\
--	&	$\gamma$ Vir 	&			37.84	 		&		0.88	 		&			0.94	 		&			0.9	 		&		0.78	 	&		0.77	 	&		135.53	 	&	5	&		\includegraphics[width=45mm]{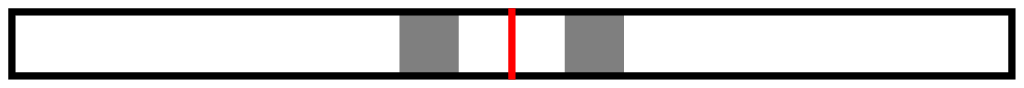} \\
--	&	$\alpha$ Com 	&			12.49	 		&		0.5	 		&			1.43	 		&			1.37	 		&		1.45	 	&		1.42	 	&		41.08	 	&	5	&		\includegraphics[width=45mm]{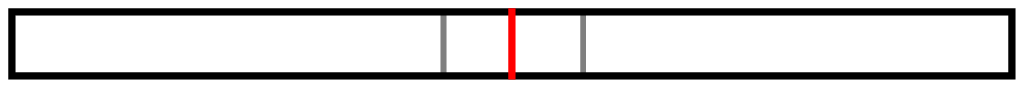} \\
--	&	$\epsilon$ CrB 	&			13.98	 		&		0.28	 		&			0.79	 		&			0.78	 		&		2.51	 	&		2.49	 	&		42.43	 	&	5	&		\includegraphics[width=45mm]{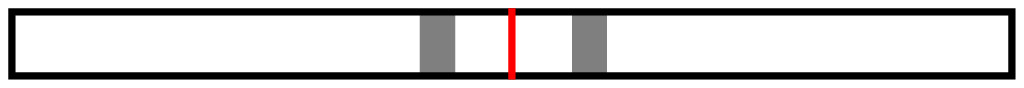} \\
--	&	BD -8$^{\rm o}$ 4352 	&			1.35	 		&		0.05	 		&			0.42	 		&			0.42	 		&		0.33	 	&		0.33	 	&		3.41	 	&	5	&		\includegraphics[width=45mm]{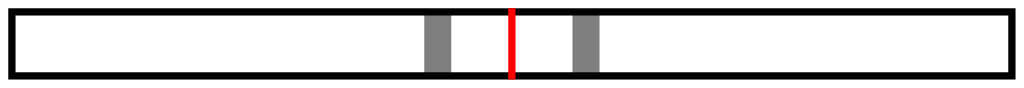} \\
--	&	BD 45$^{\rm o}$ 2505 	&			4.58	 		&		0.73	 		&			0.29	 		&			0.29	 		&		0.25	 	&		0.25	 	&		15.93	 	&	5	&		\includegraphics[width=45mm]{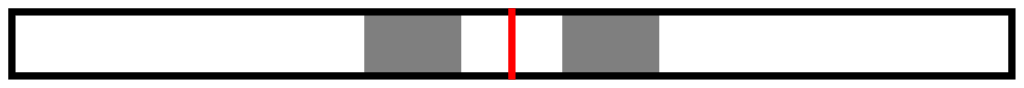} \\
\\
--	&	$\delta$ Equ 	&			4.73	 		&		0.42	 		&			1.66	 		&			1.59	 		&		0.66	 	&		0.64	 	&		15.18	 	&	5	&		\includegraphics[width=45mm]{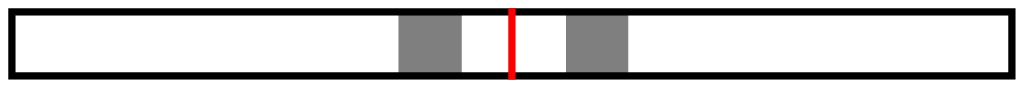} \\
--	&	Kpr 37 	 	&			9.67	 		&		0.15	 		&			1.2	 		&			0.89	 		&		2.26	 	&		1.97	 	&		27.23	 	&	5	&		\includegraphics[width=45mm]{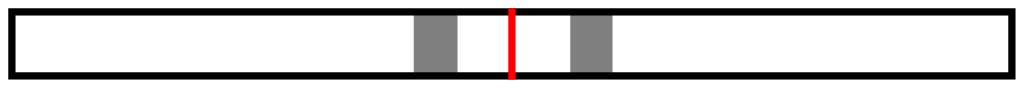} \\
--	&	99 Her 	 	&			16.39	 		&		0.74	 		&			0.89	 		&			0.52	 		&		0.98	 	&		0.77	 	&		56.97	 	&	5	&		\includegraphics[width=45mm]{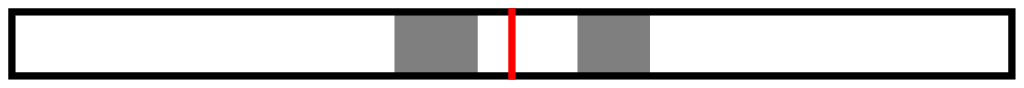} \\
--	&	9 Pup 	 	&			10.00	 		&		0.69	 		&			0.98	 		&			0.87	 		&		0.67	 	&		0.63	 	&		34.49	 	&	5	&		\includegraphics[width=45mm]{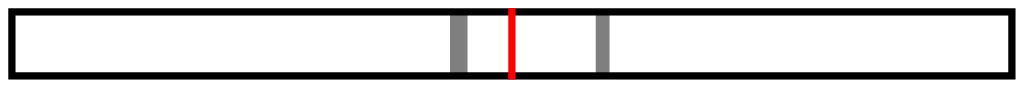} \\
--	&	$\alpha$ CMa 	&			19.89	 		&		0.59	 		&			2.11	 		&			1.04	 		&		2.12	 	&		1.54	 	&		66.70	 	&	5	&		\includegraphics[width=45mm]{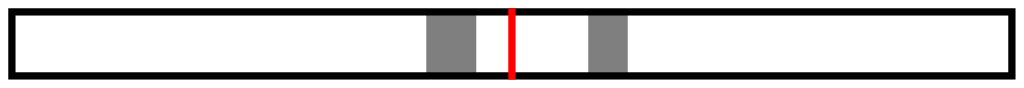} \\
\\
--	&	$\alpha$ Cen 	&			23.57	 		&		0.52	 		&			1.12	 		&			0.95	 		&		2.71	 	&		2.52	 	&		77.86	 	&	5	&		\includegraphics[width=45mm]{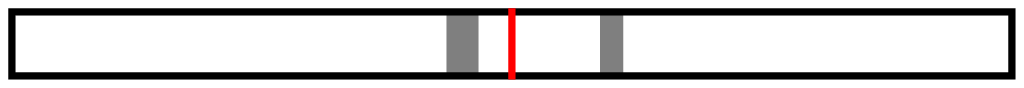} \\
--	&	$\xi$ Boo	 	&			33.14	 		&		0.51	 		&			0.86	 		&			0.73	 		&		3.85	 	&		3.58	 	&		109.35	 	&	5	&		\includegraphics[width=45mm]{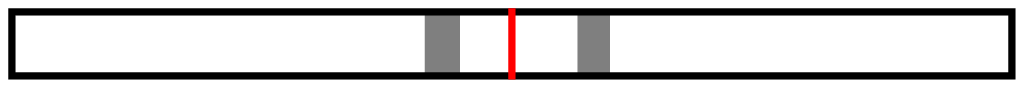} \\
--	&	G9-42 	 	&			0.44	 		&		0.81	 		&			0.77	 		&			0.04	 		&		0.02	 	&		0.01	 	&		1.45	 	&	7	&		\includegraphics[width=45mm]{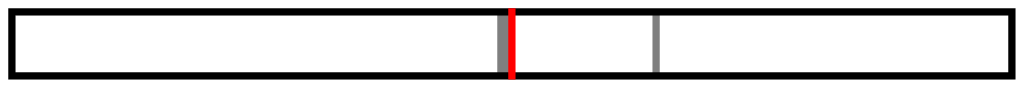} \\
--	&	G62-30 	 	&			0.79	 		&		0.59	 		&			0.68	 		&			0.04	 		&		0.1	 	&		0.03	 	&		2.49	 	&	7	&		\includegraphics[width=45mm]{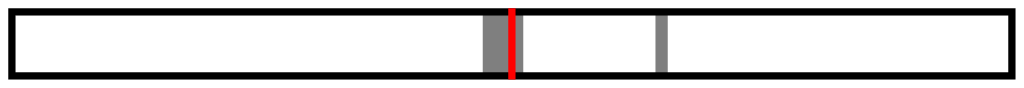} \\
--	&	G165-22 	&			0.1	 		&		0.08	 		&			0.82	 		&			0.07	 		&		0.03	 	&		0.01	 	&		0.25	 	&	7	&		\includegraphics[width=45mm]{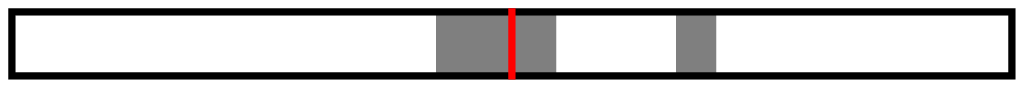} \\
\\
--	&	G65-52 	 	&			0.48	 		&		0.26	 		&			0.62	 		&			0.04	 		&		0.12	 	&		0.04	 	&		1.36	 	&	7	&		\includegraphics[width=45mm]{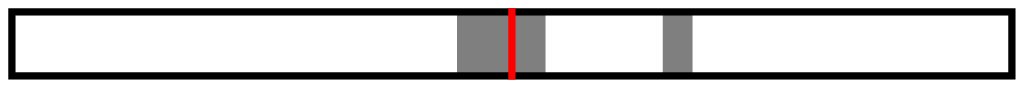} \\
--	&	G178-27 	&			0.33	 		&		0.43	 		&			0.68	 		&			0.05	 		&		0.06	 	&		0.02	 	&		1	 	&	7	&		\includegraphics[width=45mm]{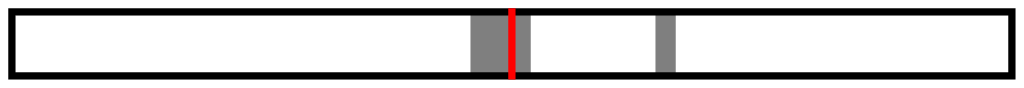} \\
--	&	G15-6 	 	&			0.73	 		&		0.45	 		&			0.67	 		&			0.06	 		&		0.13	 	&		0.04	 	&		2.25	 	&	7	&		\includegraphics[width=45mm]{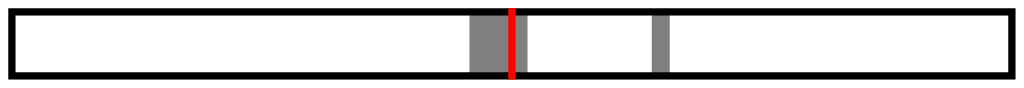} \\
--	&	G66-65 	 	&			0.61	 		&		0.9	 		&			0.7	 		&			0.04	 		&		0.01	 	&		0	 	&		2.05	 	&	7	&		\includegraphics[width=45mm]{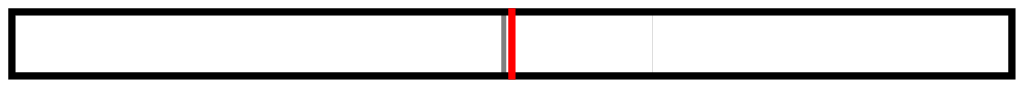} \\
--	&	G141-8 	 	&			0.8	 		&		0.58	 		&			0.77	 		&			0.02	 		&		0.11	 	&		0.02	 	&		2.43	 	&	7	&		\includegraphics[width=45mm]{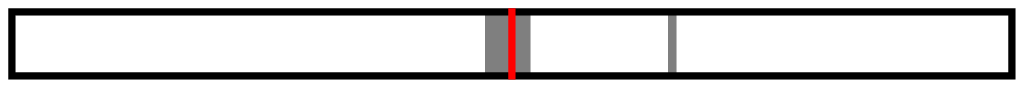} \\

\hline
\end{tabular}
\end{figure*}
\end{landscape}

\begin{landscape}
\begin{figure*}
\begin{tabular}{llllllllllc}
\hline
Object		&Alter. name&	 $a$	&      $e$	&    	$M_{1}$		&  	$M_{2}$		&	$R{ce1}$	&	$R{ce2}$	&	$R_{cb}$	&Ref&	Scheme	\\
 HIP		&	&       (AU)	&		&	(M$_{\odot}$)		&	(M$_{\odot}$)		&	 (AU)		&  	(AU)		&  	(AU)		& &  	 $*_{1}$  \   $*_{2}$ \\
\hline

--	&	G18-35 			&			4.48	 		&		0.37			&			0.75		 	&			0.07	 		&		0.93	 	&		0.32	 	&		13.44	 	&	7	&		\includegraphics[width=45mm]{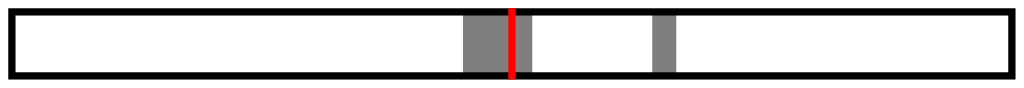} \\

--	&	V821 Cas		&			0.044			&		0.13			&			2.046			&			1.626			&		0.01		&		0.01		&		0.12		&	10	&		\includegraphics[width=45mm]{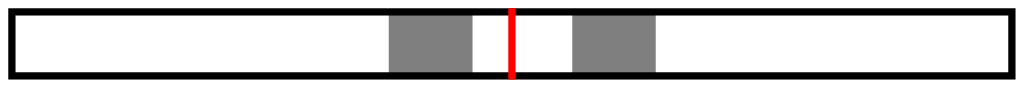} \\
--	&	SV Cam			&			0.016			&		0			&			0.862			&			0.646			&		0.0045		&		0.004		&		0.03		&	11	&		\includegraphics[width=45mm]{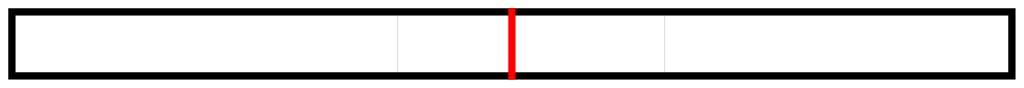} \\
  --	&	BS Dra			&			0.060			&		0			&			1.294			&			1.276			&		0.02		&		0.016		&		0.11		&	11	&		\includegraphics[width=45mm]{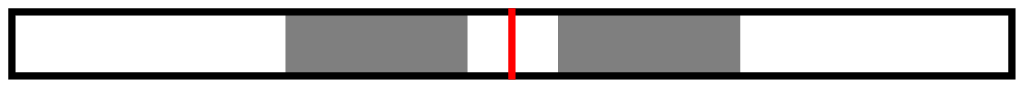} \\
473	&	HD 38			&			73.01		 	&		0.22			&			0.76	 		&			0.74	 		&		14.41	 	&		14.24	 	&		215.35	 	&	3	&		\includegraphics[width=45mm]{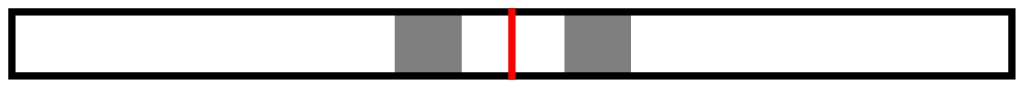} \\
\\
1349	&	HD 1273	&			1.25	 		&		0.57	 		&			0.98	 		&			0.55	 		&		0.13	 	&		0.1	 	&		4.18	 	&	1	&	\includegraphics[width=45mm]{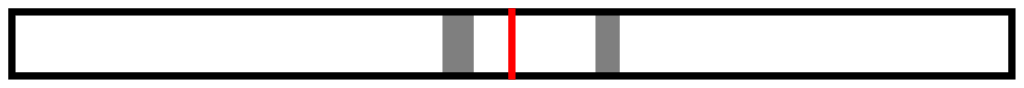} \\
1955	&	HD 2070	&			0.54	 		&		0.33	 		&			1.13	 		&			0.48	 		&		0.1	 	&		0.07	 	&		1.66	 	&	1	&		\includegraphics[width=45mm]{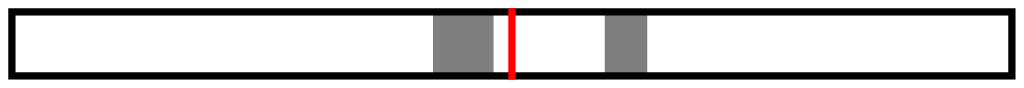} \\
2237	&	HD 2475	&			7.08	 		&		0	 		&			1.56	 		&			1.24	 		&		1.96	 	&		1.76	 	&		12.87	 	&	2	&		\includegraphics[width=45mm]{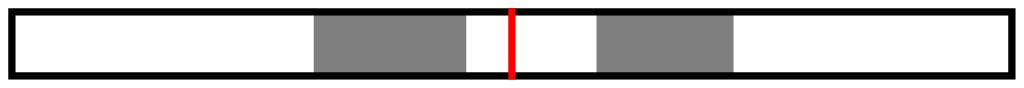} \\
2552	&		&			41.32	 		&		0	 		&			0.4	 		&			0.29	 		&		11.6	 	&		10.02	 	&		75.05	 	&	3	&		\includegraphics[width=45mm]{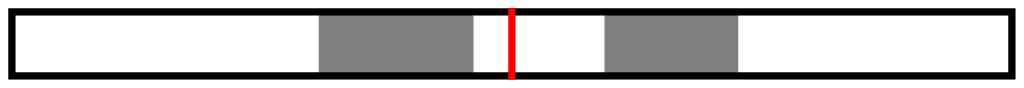} \\
2762	&	HD 3196	&			5.13	 		&		0.77	 		&			1.71	 		&			1.14	 		&		0.25	 	&		0.21	 	&		17.96	 	&	2	&		\includegraphics[width=45mm]{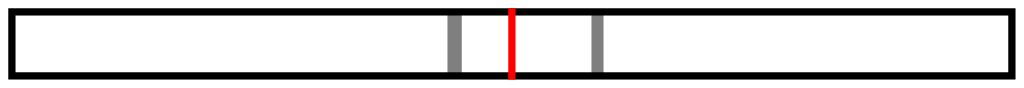} \\
\\
2941	&	ADS520 	 	&			9.57	 		&		0.22	 		&			0.7	 		&			0.7	 		&		1.87	 	&		1.87	 	&		28.23	 	&	5	&		\includegraphics[width=45mm]{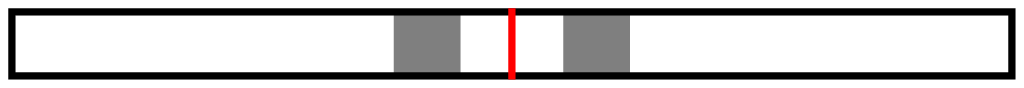} \\
3821	&	HD 4614 	&			72	 		&		0.49	 		&			0.99	 		&			0.51	 		&		9.54	 	&		7.05	 	&		235.07	 	&	4	&		\includegraphics[width=45mm]{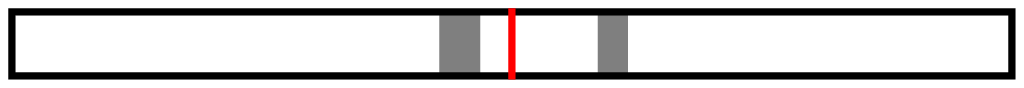} \\
3850	&	HD 4747   	&			6.7	 		&		0.64	 		&			0.82	 		&			0.04	 		&		0.73	 	&		0.19	 	&		21.21	 	&	4	&		\includegraphics[width=45mm]{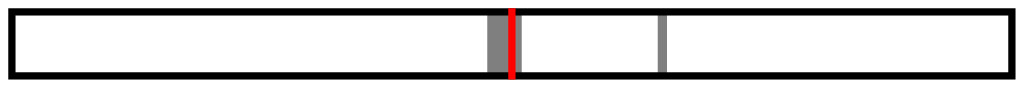} \\
5531	&			&			5.00			&		0.72			&			1.17			&			1.16			&		0.29		&		0.29		&		17.36		&	8	&		\includegraphics[width=45mm]{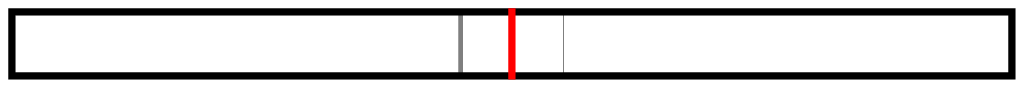} \\
5842	&	HD 7693   	&			23.4	 		&		0.04	 		&			0.89	 		&			0.84	 		&		5.96	 	&		5.81	 	&		57.89	 	&	4	&		\includegraphics[width=45mm]{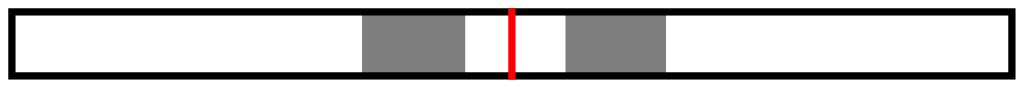} \\
\\
7078	&	HD 9021	&			0.64	 		&		0.31	 		&			1.21	 		&			0.7	 		&		0.12	 	&		0.09	 	&		1.97	 	&	1	&		\includegraphics[width=45mm]{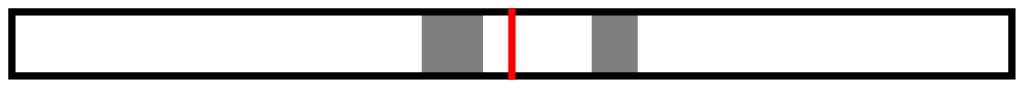} \\
7751	&	HD 10360   	&			52.2	 		&		0.53	 		&			0.77	 		&			0.75	 		&		5.61	 	&		5.54	 	&		173.15	 	&	4	&		\includegraphics[width=45mm]{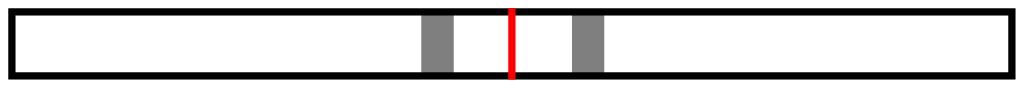} \\
7918	&	HD 10307	&			7.1	 		&		0.42	 		&			0.8	 		&			0.14	 		&		1.26	 	&		0.57	 	&		22.13	 	&	2	&		\includegraphics[width=45mm]{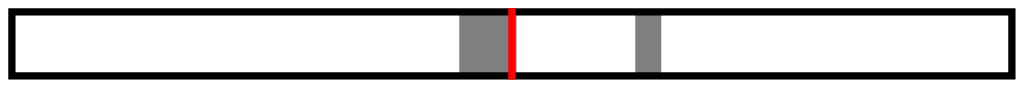} \\
8903	&	HD 11636	&			0.63	 		&		0.88	 		&			1.86	 		&			1.05	 		&		0.01	 	&		0.01	 	&		2.25	 	&	1	&		\includegraphics[width=45mm]{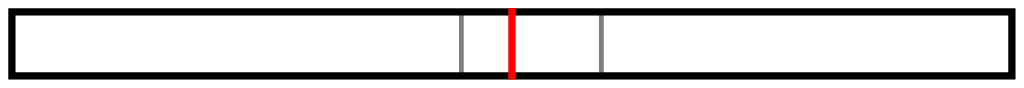} \\
8903	&	HD 11636	&			0.66	 		&		0.9	 		&			2.07	 		&			1.28	 		&		0.01	 	&		0.01	 	&		2.37	 	&	2	&		\includegraphics[width=45mm]{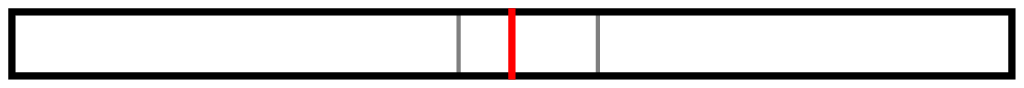} \\
\\
48904	&			&			0.027			&		0.01			&			0.365			&			0.348			&		0.007		&		0.006		&		0.06		&	9	&		\includegraphics[width=45mm]{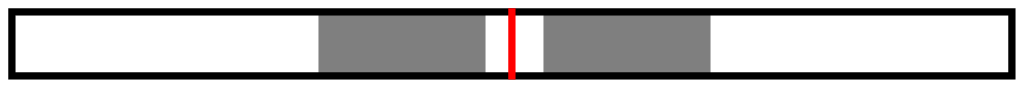} \\
9480	&	WDS 02019+7054	&			22.5	 		&		0.39	 		&			1.92	 		&			1.19	 		&		3.59	 	&		2.88	 	&		71.31	 	&	6	&		\includegraphics[width=45mm]{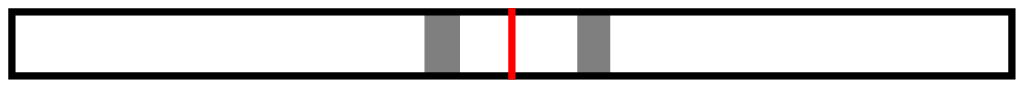} \\
\bf{10138}&	\bf{HD 13445}	&			\bf{18.4}		&		\bf{0.4}		&			\bf{0.77}		&			\bf{0.49}		&		\bf{2.86}	&		\bf{2.33}	&		\bf{58.53}	&	\bf{13}	&		\includegraphics[width=45mm]{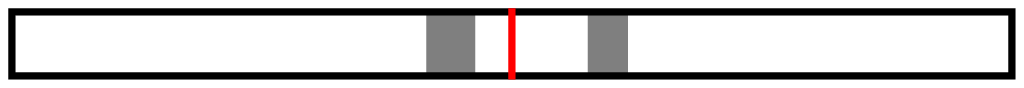} \\
10321	&	HD 13507   	&			4.3	 		&		0.14	 		&			1	 		&			0.05	 		&		1.34	 	&		0.36	 	&		11.18	 	&	4	&		\includegraphics[width=45mm]{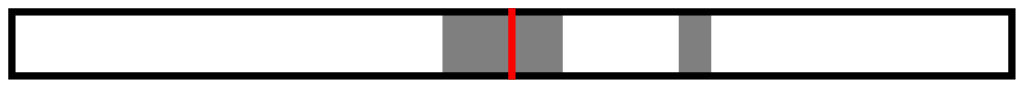} \\
11231	&	HD 15064	&			0.64	 		&		0.29	 		&			1.01	 		&			0.68	 		&		0.12	 	&		0.1	 	&		1.95	 	&	1	&		\includegraphics[width=45mm]{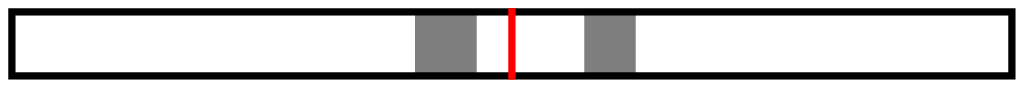} \\

\hline
\end{tabular}
\end{figure*}
\end{landscape}

\begin{landscape}
\begin{figure*}
\begin{tabular}{llllllllllc}
\hline
Object		&Alter. name&	 $a$	&      $e$	&    	$M_{1}$		&  	$M_{2}$		&	$R{ce1}$	&	$R{ce2}$	&	$R_{cb}$	&Ref&	Scheme	\\
 HIP		&	&       (AU)	&		&	(M$_{\odot}$)		&	(M$_{\odot}$)		&	 (AU)		&  	(AU)		&  	(AU)		& &  	 $*_{1}$  \   $*_{2}$ \\
\hline
12062	&	HD 15862	&			2.04	 		&		0.26	 		&			0.95	 		&			0.44	 		&		0.43	 	&		0.3	 	&		6.11	 	&	1	&		\includegraphics[width=45mm]{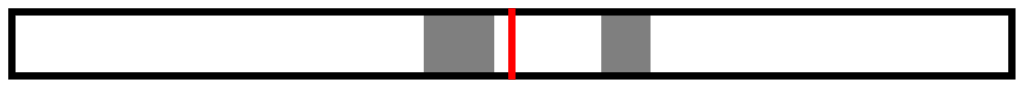} \\
12114	&	HD 16160   	&			15	 		&		0.75	 		&			0.76	 		&			0.09	 		&		1.01	 	&		0.39	 	&		50.26	 	&	4	&		\includegraphics[width=45mm]{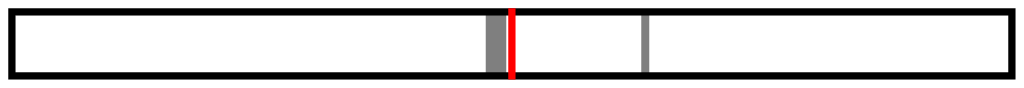} \\
12153	&	HD 16234	&			4.22	 		&		0.88	 		&			11	 		&			9.41	 		&		0.09	 	&		0.08	 	&		15.11	 	&	2	&		\includegraphics[width=45mm]{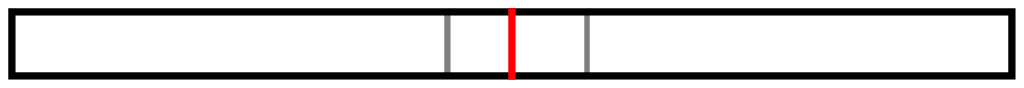} \\
12623	&	HD 16739	&			1.27	 		&		0.66	 		&			1.13	 		&			1.39	 		&		0.1	 	&		0.09	 	&		4.35	 	&	2	&		\includegraphics[width=45mm]{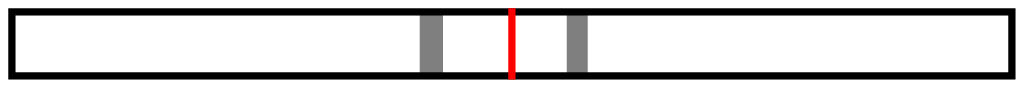} \\
12777	&	HD 16895   	&			249.5	 		&		0.13	 		&			1.24	 		&			0.43	 		&		66.5	 	&		41.09	 	&		684.25	 	&	4	&		\includegraphics[width=45mm]{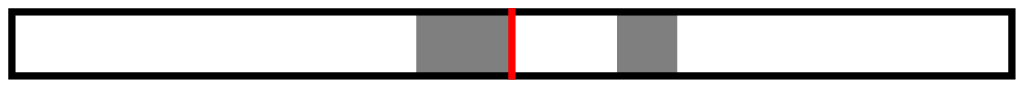} \\
\\
13769	&	HD 18445   	&			1.06	 		&		0.56	 		&			0.78	 		&			0.18	 		&		0.13	 	&		0.07	 	&		3.47	 	&	4	&		\includegraphics[width=45mm]{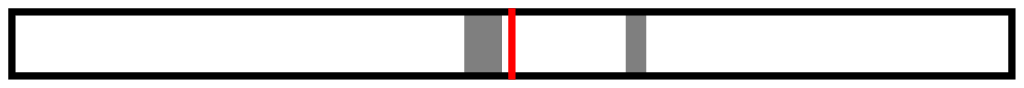} \\
\bf{14954}	&	\bf{HD 19994}	&	\bf{120}	&	\bf{0.26}	&	\bf{1.35}	&	\bf{0.35}	&	\bf{27.34}	&	\bf{14.83}	&	\bf{354.86}	&	\bf{13}	&		\includegraphics[width=45mm]{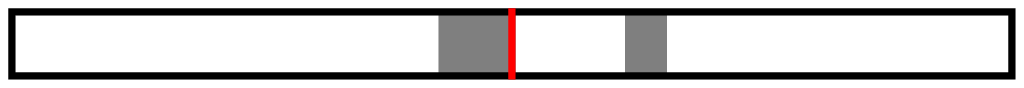} \\
18512	&	HD 24916	&			174.55	 		&		0	 		&			0.35	 		&			0.17	 		&		52.36	 	&		37.68	 	&		315.65	 	&	3	&		\includegraphics[width=45mm]{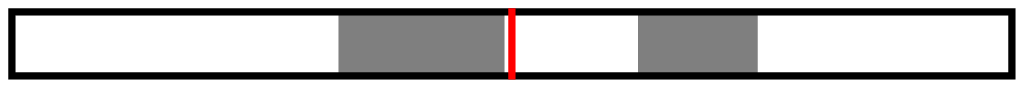} \\
19206	&		&	9.290	&	0.69	&	0.960	&	0.790	&	0.64	&	0.58	&	32.1	&	8	&		\includegraphics[width=45mm]{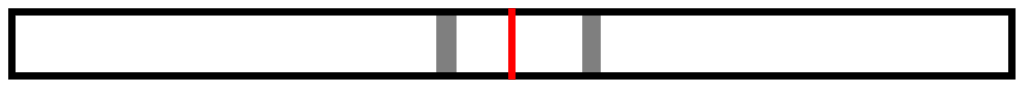} \\
20087	&	HD 27176	&			7.05	 		&		0.17	 		&			1.76	 		&			0.95	 		&		1.66	 	&		1.26	 	&		20.08	 	&	2	&		\includegraphics[width=45mm]{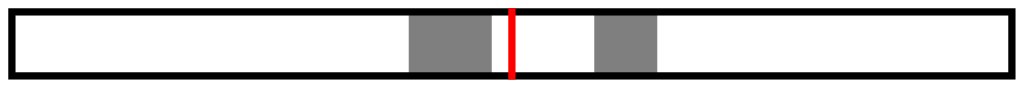} \\
\\
20935	&	HD 28394	&			0.99	 		&		0.24	 		&			1.13	 		&			1.11	 		&		0.19	 	&		0.19	 	&		2.95	 	&	1	&		\includegraphics[width=45mm]{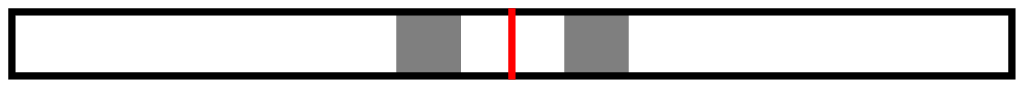} \\
22429	&	HD 30339   	&			0.13	 		&		0.25	 		&			1.39	 		&			0.07	 		&		0.03	 	&		0.01	 	&		0.36	 	&	4	&		\includegraphics[width=45mm]{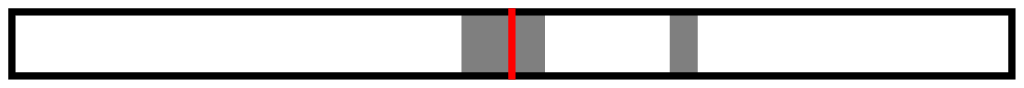} \\
23395	&	WDS 05017+2640 	&			10.28	 		&		0.33	 		&			1	 		&			0.72	 		&		1.79	 	&		1.54	 	&		31.9	 	&	6	&		\includegraphics[width=45mm]{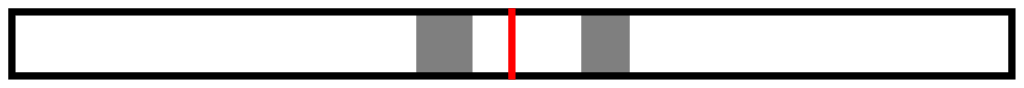} \\
23835	&	HD 32923   	&			2.86	 		&		0.9	 		&			1.11	 		&			1.03	 		&		0.05	 	&		0.05	 	&		10.28	 	&	4	&		\includegraphics[width=45mm]{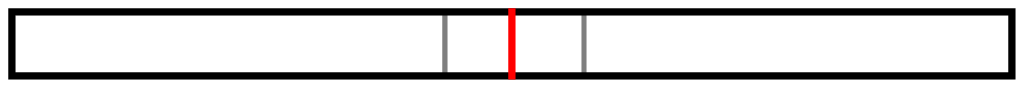} \\
24419	&	HD 34101	&			1.75	 		&		0.08	 		&			0.9	 		&			0.21	 		&		0.52	 	&		0.27	 	&		4.52	 	&	1	&		\includegraphics[width=45mm]{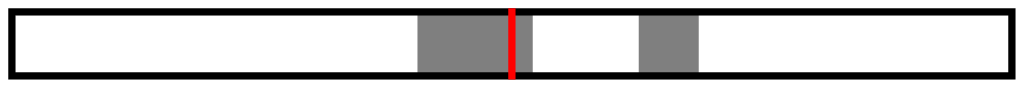} \\
\\
25662	&	HD 35956   	&			2.6	 		&		0.62	 		&			0.98	 		&			0.18	 		&		0.28	 	&		0.13	 	&		8.58	 	&	4	&		\includegraphics[width=45mm]{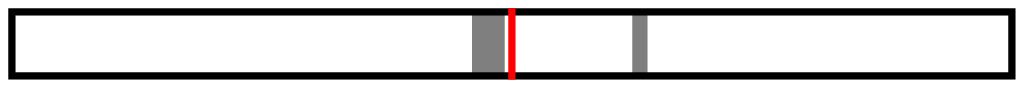} \\
27913	&	HD 39587   	&			5.9	 		&		0.45	 		&			1.05	 		&			0.14	 		&		1.01	 	&		0.41	 	&		18.41	 	&	4	&		\includegraphics[width=45mm]{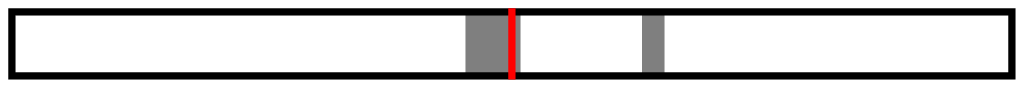} \\
29860	&	HD 43587   	&			11.6	 		&		0.8	 		&			1.06	 		&			0.34	 		&		0.54	 	&		0.32	 	&		40.39	 	&	4	&		\includegraphics[width=45mm]{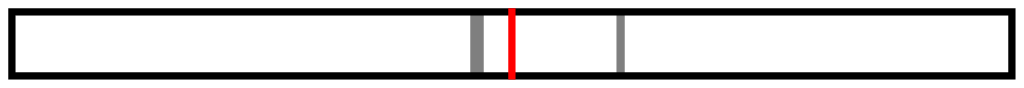} \\
33451	&	HD 51825	&			9.3	 		&		0.43	 		&			1.61	 		&			1.26	 		&		1.31	 	&		1.17	 	&		29.93	 	&	2	&		\includegraphics[width=45mm]{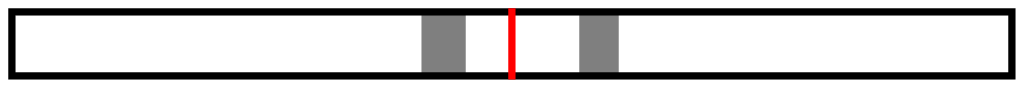} \\
34164	&	HD 53424	&			1.7	 		&		0.27	 		&			1.09	 		&			0.66	 		&		0.34	 	&		0.27	 	&		5.13	 	&	1	&		\includegraphics[width=45mm]{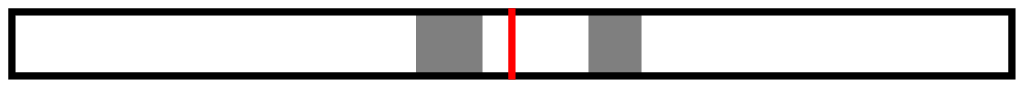} \\
\\
38657	&	HD 64468   	&			0.56	 		&		0.26	 		&			0.81	 		&			0.14	 		&		0.13	 	&		0.06	 	&		1.64	 	&	4	&		\includegraphics[width=45mm]{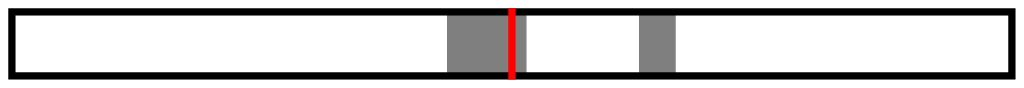} \\
39064	&	HD 65430   	&			4	 		&		0.32	 		&			0.83	 		&			0.06	 		&		0.92	 	&		0.29	 	&		11.66	 	&	4	&		\includegraphics[width=45mm]{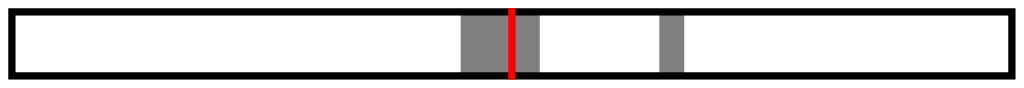} \\
39893	&		&			1.81	 		&		0.21	 		&			0.95	 		&			0.52	 		&		0.4	 	&		0.31	 	&		5.29	 	&	1	&		\includegraphics[width=45mm]{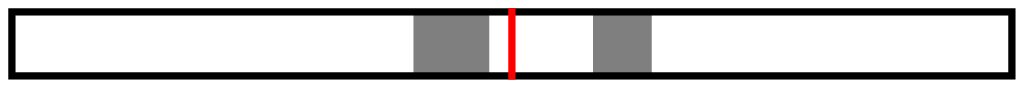} \\
44248	&	HD 76943	&			10.51	 		&		0.1	 		&			1.53	 		&			0.92	 		&		2.69	 	&		2.13	 	&		28.27	 	&	2	&		\includegraphics[width=45mm]{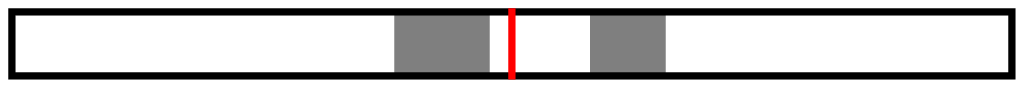} \\
44892	&	HD 78418	&	0.184	&	0.2	&	1.173	&	1.011	&	0.04	&	0.04	&	0.53	&	12	&		\includegraphics[width=45mm]{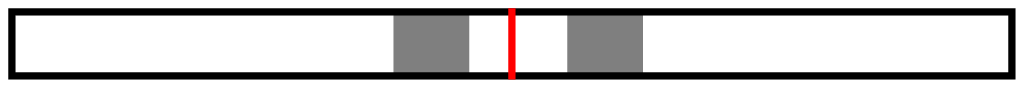} \\

\hline
\end{tabular}
\end{figure*}
\end{landscape}

\begin{landscape}
\begin{figure*}
\begin{tabular}{llllllllllc}
\hline
Object		&Alter. name&	 $a$	&      $e$	&    	$M_{1}$		&  	$M_{2}$		&	$R{ce1}$	&	$R{ce2}$	&	$R_{cb}$	&Ref&	Scheme	\\
 HIP		&	&       (AU)	&		&	(M$_{\odot}$)		&	(M$_{\odot}$)		&	 (AU)		&  	(AU)		&  	(AU)		& &  	 $*_{1}$  \   $*_{2}$ \\
\hline
45571	&	HD 80671	&			4.22	 		&		0.51	 		&			3.66	 		&			3.66	 		&		0.47	 	&		0.47	 	&		13.92	 	&	2	&		\includegraphics[width=45mm]{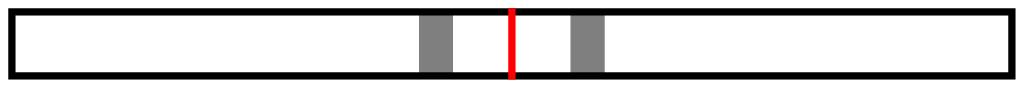} \\
52940	&	HI52940 		&			2.6	 		&		0.37	 		&			1.12	 		&			0.12	 		&		0.53	 	&		0.2	 	&		7.84	 	&	4	&		\includegraphics[width=45mm]{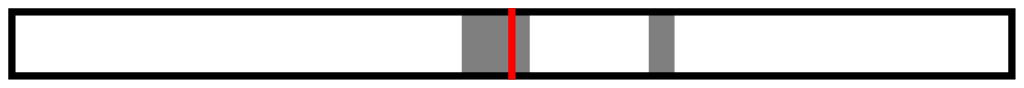} \\
54204	&	WDS 11053-2718 	&			6.04	 		&		0.35	 		&			1.93	 		&			1.93	 		&		0.95	 	&		0.95	 	&		18.91	 	&	6	&		\includegraphics[width=45mm]{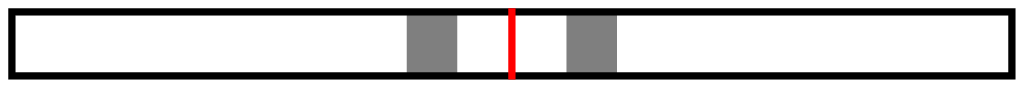} \\
55016	&	HD 97907	&			6.87	 		&		0.42	 		&			2.62	 		&			2.32	 		&		0.97	 	&		0.92	 	&		22.05	 	&	2	&		\includegraphics[width=45mm]{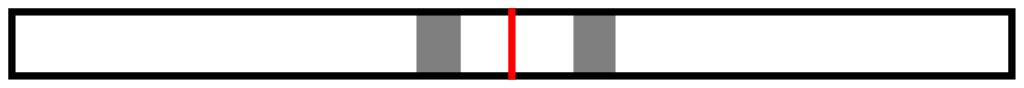} \\
56809	&	HD 101177	&			240.39	 		&		0.05	 		&			1.95	 		&			1.36	 		&		63.9	 	&		54.21	 	&		605.53	 	&	3	&		\includegraphics[width=45mm]{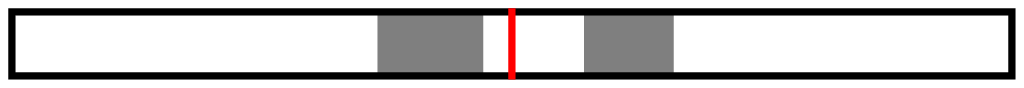} \\
\\
60129	&	HD 107259	&			10.48	 		&		0.08	 		&			6.01	 		&			0.67	 		&		3.37	 	&		1.26	 	&		26.45	 	&	2	&		\includegraphics[width=45mm]{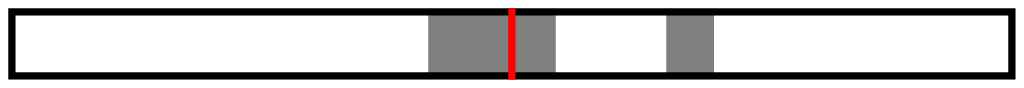} \\
63406	&	HD 112914	&			1.59	 		&		0.33	 		&			0.82	 		&			0.23	 		&		0.32	 	&		0.18	 	&		4.86	 	&	1	&		\includegraphics[width=45mm]{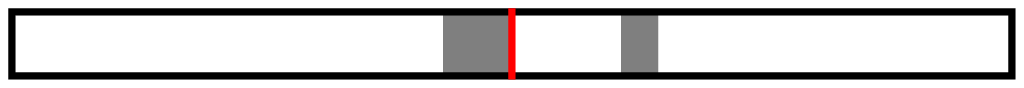} \\
65026	&	WDS 13198+4747 	&			14.36	 		&		0.23	 		&			0.66	 		&			0.58	 		&		2.85	 	&		2.68	 	&		42.58	 	&	6	&		\includegraphics[width=45mm]{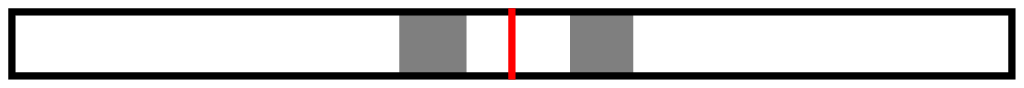} \\
65343	&	HD 116495	&			39.65	 		&		0.84	 		&			0.61	 		&			0.58	 		&		1.17	 	&		1.15	 	&		140.96	 	&	3	&		\includegraphics[width=45mm]{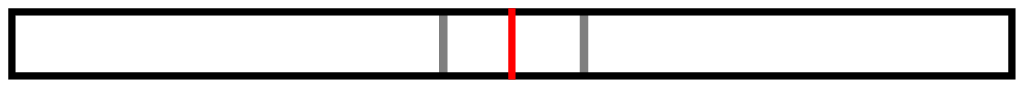} \\
66077	&		&			51.51	 		&		0	 		&			0.39	 		&			0.35	 		&		13.91	 	&		13.24	 	&		93.65	 	&	3	&		\includegraphics[width=45mm]{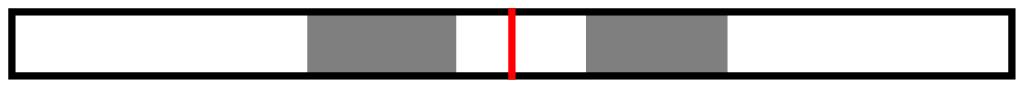} \\
\\
66492	&		&			46.59	 		&		0.61	 		&			0.54	 		&			0.39	 		&		4.23	 	&		3.65	 	&		157.58	 	&	3	&		\includegraphics[width=45mm]{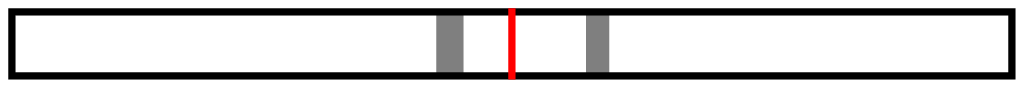} \\
66640	&	WDS 13396+1044	&			10.7	 		&		0.55	 		&			1.24	 		&			1.19	 		&		1.09	 	&		1.07	 	&		35.68	 	&	6	&		\includegraphics[width=45mm]{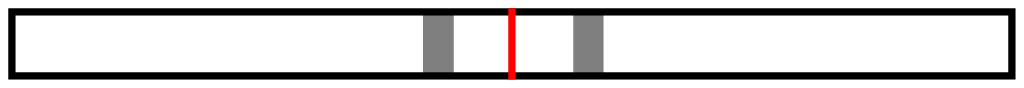} \\
\bf{67275}	&	\bf{HD 120136}	&	\bf{245}	&	\bf{0.91}	&	\bf{1.35}	&	\bf{0.4}	&	\bf{4.38}	&	\bf{2.52}	&	\bf{868.91}	&	\bf{13}	&		\includegraphics[width=45mm]{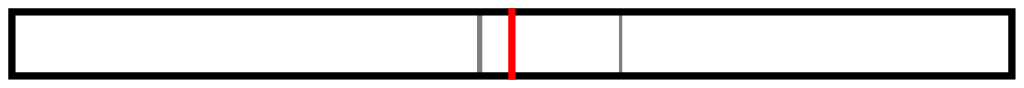} \\
67422	&	HD 120476   	&			33.2	 		&		0.45	 		&			0.76	 		&			0.75	 		&		4.3	 	&		4.27	 	&		107.59	 	&	3	&		\includegraphics[width=45mm]{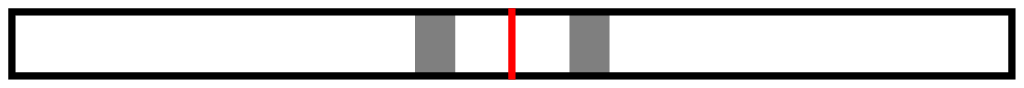} \\
67422	&	HD 120476   	&			33.15	 		&		0.44	 		&			0.83	 		&			0.76	 		&		4.45	 	&		4.27	 	&		107.08	 	&	4	&		\includegraphics[width=45mm]{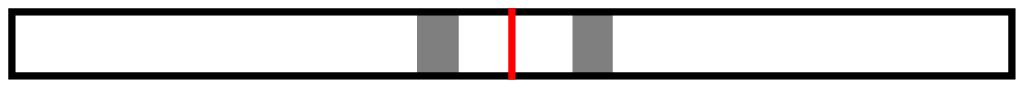} \\
\\
68682	&	HD 122742   	&			5.46	 		&		0.55	 		&			1.11	 		&			0.55	 		&		0.63	 	&		0.45	 	&		18.11	 	&	2	&		\includegraphics[width=45mm]{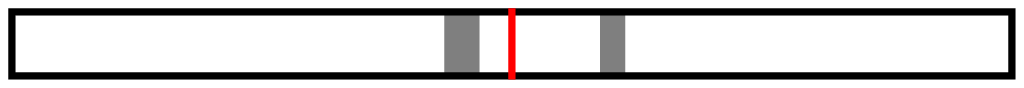} \\
68682	&	HD 122742   	&			5.3	 		&		0.48	 		&			0.92	 		&			0.54	 		&		0.7	 	&		0.55	 	&		17.28	 	&	4	&		\includegraphics[width=45mm]{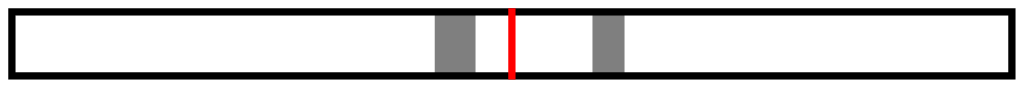} \\
69226	&	HD 123999	&	0.124	&	0.19	&	1.411	&	1.368	&	0.026	&	0.025	&	0.359	&	12	&		\includegraphics[width=45mm]{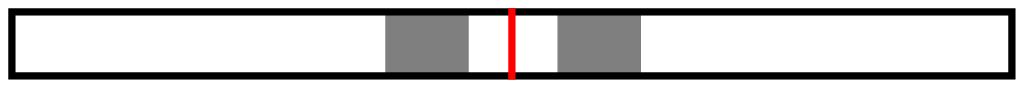} \\
71094	&		&			13.98	 		&		0.16	 		&			1.89	 		&			1.16	 		&		3.28	 	&		2.62	 	&		39.6	 	&	2	&		\includegraphics[width=45mm]{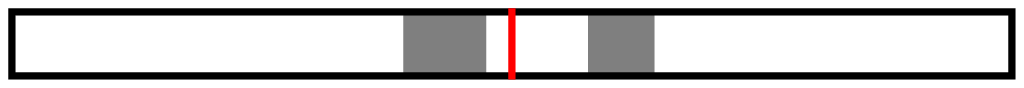} \\
71681	&	HD 128621   	&			22.76	 		&		0.51	 		&			1.12	 		&			0.89	 		&		2.67	 	&		2.4	 	&		75.04	 	&	4	&		\includegraphics[width=45mm]{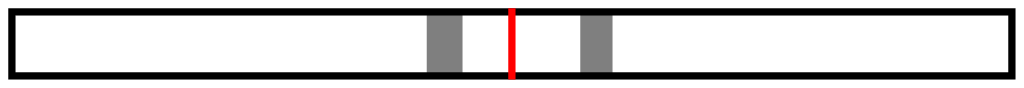} \\
\\
71683	&	HD 128620   	&			22.76	 		&		0.51	 		&			1.12	 		&			0.89	 		&		2.67	 	&		2.4	 	&		75.04	 	&	4	&		\includegraphics[width=45mm]{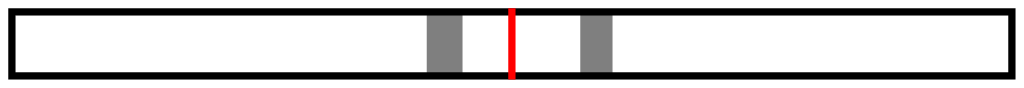} \\
71729	&	HD 129132	&			8.28	 		&		0.4	 		&			3.34	 		&			3.29	 		&		1.19	 	&		1.18	 	&		26.4	 	&	2	&		\includegraphics[width=45mm]{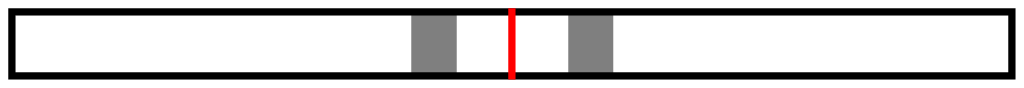} \\
72659	&	HD 131156   	&			32.8	 		&		0.51	 		&			0.92	 		&			0.79	 		&		3.8	 	&		3.54	 	&		108.17	 	&	4	&		\includegraphics[width=45mm]{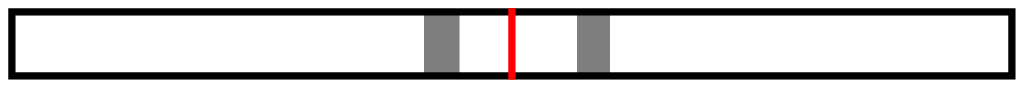} \\
72848	&	HD 131511   	&			0.53	 		&		0.51	 		&			0.79	 		&			0.45	 		&		0.07	 	&		0.05	 	&		1.74	 	&	1	&		\includegraphics[width=45mm]{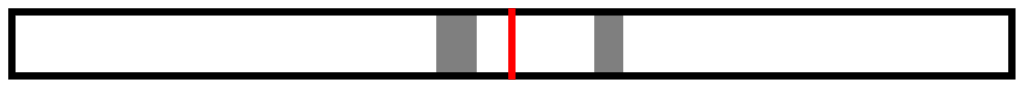} \\
72848	&	HD 131511   	&			0.52	 		&		0.51	 		&			0.93	 		&			0.45	 		&		0.07	 	&		0.05	 	&		1.71	 	&	4	&		\includegraphics[width=45mm]{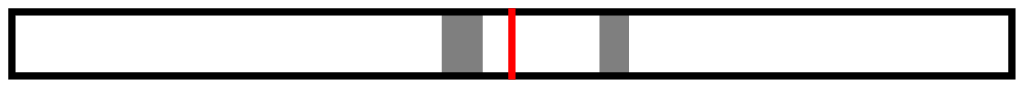} \\

\hline
\end{tabular}
\end{figure*}
\end{landscape}

\begin{landscape}
\begin{figure*}
\begin{tabular}{llllllllllc}
\hline
Object		&Alter. name&	 $a$	&      $e$	&    	$M_{1}$		&  	$M_{2}$		&	$R{ce1}$	&	$R{ce2}$	&	$R_{cb}$	&Ref&	Scheme	\\
 HIP		&	&       (AU)	&		&	(M$_{\odot}$)		&	(M$_{\odot}$)		&	 (AU)		&  	(AU)		&  	(AU)		& &  	 $*_{1}$  \   $*_{2}$ \\
\hline
73440	&	HD 133621	&			1.25	 		&		0.22	 		&			1.03	 		&			0.15	 		&		0.32	 	&		0.14	 	&		3.56	 	&	1	&		\includegraphics[width=45mm]{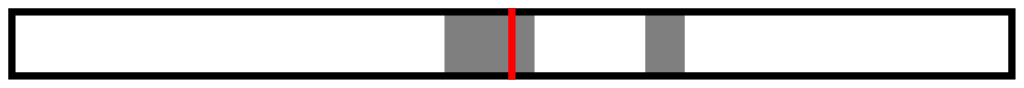} \\
74392	&	WDS 15123+1947	&			14.93	 		&		0.25	 		&			3.54	 		&			2.51	 		&		2.98	 	&		2.55	 	&		44.69	 	&	6	&		\includegraphics[width=45mm]{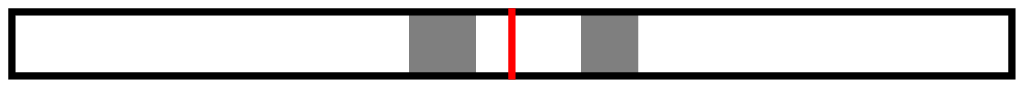} \\
75312	&	WDS 15232+3018 	&			16.15	 		&		0.26	 		&			1.26	 		&			1.18	 		&		3.02	 	&		2.93	 	&		48.64	 	&	6	&		\includegraphics[width=45mm]{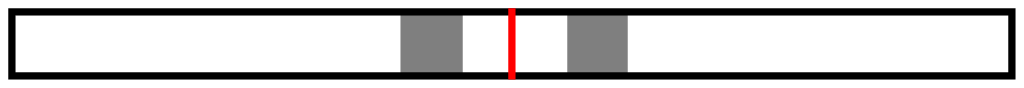} \\
75379	&	HD 137502	&			0.91	 		&		0.68	 		&			1.26	 		&			0.68	 		&		0.07	 	&		0.05	 	&		3.12	 	&	1	&		\includegraphics[width=45mm]{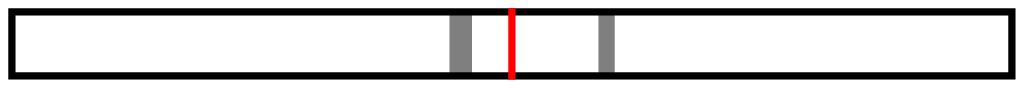} \\
76382	&	ADS9716 		&		19.15	 		&		0.59	 		&			1.14	 		&			1.14	 		&		1.73	 	&		1.73	 	&		64.54	 	&	5	&		\includegraphics[width=45mm]{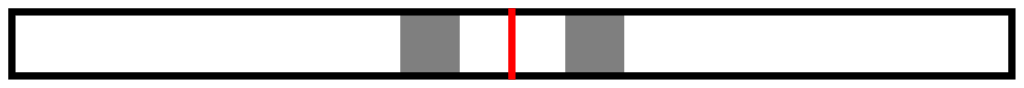} \\
\\
76852	&	HD 140159	&			12.4	 		&		0.15	 		&			2	 		&			1.98	 		&		2.7	 	&		2.69	 	&		34.96	 	&	2	&		\includegraphics[width=45mm]{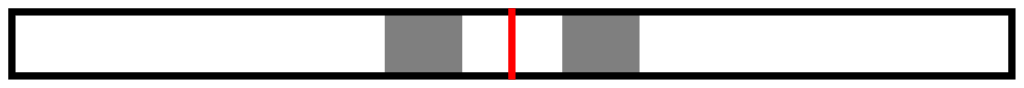} \\
77152	&	HD 140913   	&			0.55	 		&		0.54	 		&			1.17	 		&			0.04	 		&		0.08	 	&		0.02	 	&		1.67	 	&	4	&		\includegraphics[width=45mm]{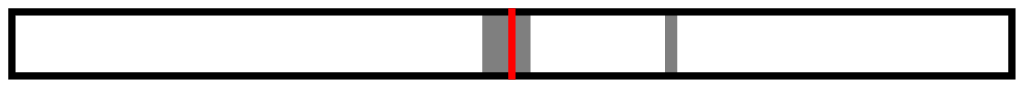} \\
78727	&	WDS 16044+1122	&			15.64	 		&		0.71	 		&			0.92	 		&			0.92	 		&		0.94	 	&		0.94	 	&		54.18	 	&	6	&		\includegraphics[width=45mm]{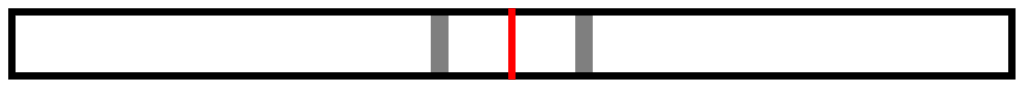} \\
79101	&	HD 145389	&			2.24	 		&		0.47	 		&			3.47	 		&			1.31	 		&		0.33	 	&		0.21	 	&		7.23	 	&	1	&		\includegraphics[width=45mm]{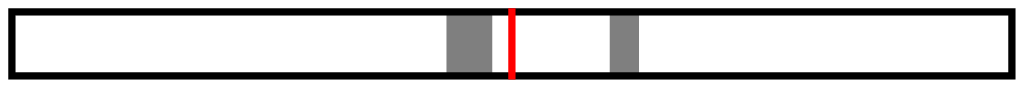} \\
80346	&		&			2.07	 		&		0.67	 		&			0.5	 		&			0.13	 		&		0.18	 	&		0.1	 	&		6.98	 	&	1	&		\includegraphics[width=45mm]{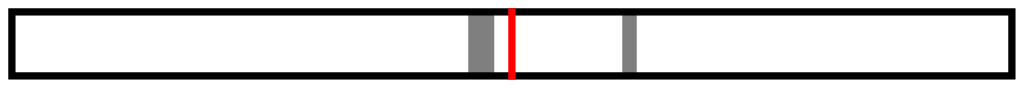} \\
\\
80686	&	HD 147584	&			0.12	 		&		0.06	 		&			1.05	 		&			0.37	 		&		0.04	 	&		0.02	 	&		0.3	 	&	1	&		\includegraphics[width=45mm]{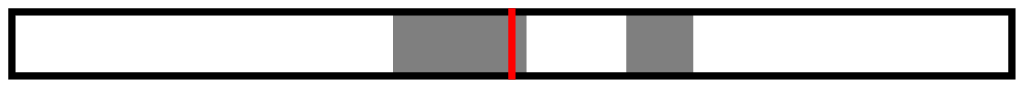} \\
81126	&	HD 149630	&			6.33	 		&		0.53	 		&			3.04	 		&			1.5	 		&		0.77	 	&		0.55	 	&		20.89	 	&	2	&		\includegraphics[width=45mm]{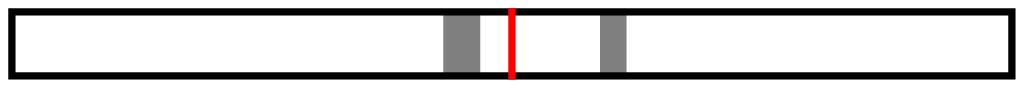} \\
82817	&	HD 152771	&			1.38	 		&		0.05	 		&			0.33	 		&			0.56	 		&		0.38	 	&		0.3	 	&		3.47	 	&	2	&		\includegraphics[width=45mm]{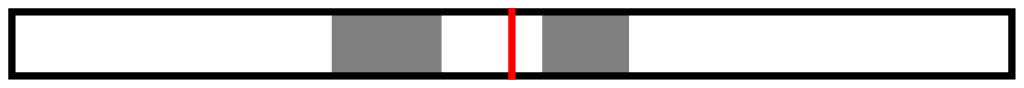} \\
82860	&	HD 153597	&			0.33	 		&		0.21	 		&			1.18	 		&			0.52	 		&		0.08	 	&		0.05	 	&		0.96	 	&	1	&		\includegraphics[width=45mm]{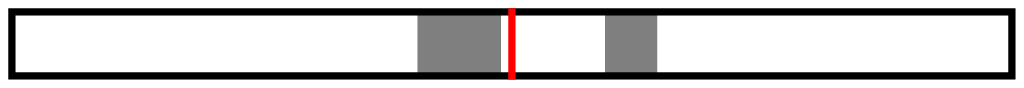} \\
83895	&	HD 155763	&			7.09	 		&		0	 		&			5.94	 		&			3.65	 		&		2.05	 	&		1.64	 	&		12.86	 	&	2	&		\includegraphics[width=45mm]{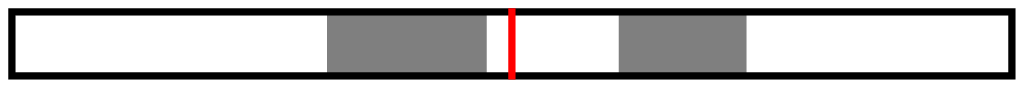} \\
\\
84140	&	HD 155876	&			5.01	 		&		0.75	 		&			0.38	 		&			0.37	 		&		0.25	 	&		0.25	 	&		17.5	 	&	2	&		\includegraphics[width=45mm]{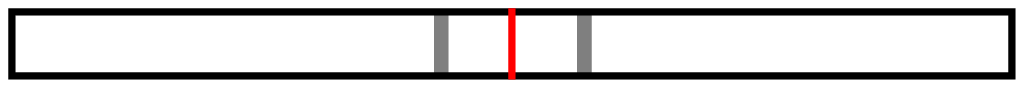} \\
84720	&	HD 156274   	&			91.65	 		&		0.78	 		&			0.79	 		&			0.47	 		&		4.33	 	&		3.41	 	&		321.18	 	&	4	&		\includegraphics[width=45mm]{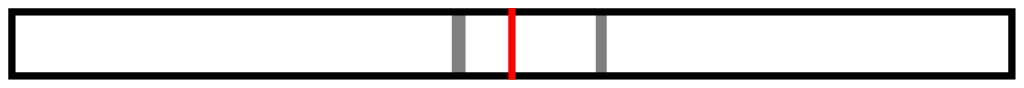} \\
84949	&	HD 157482	&			4.87	 		&		0.67	 		&			1.15	 		&			2.62	 		&		0.29	 	&		0.42	 	&		16.6	 	&	2	&		\includegraphics[width=45mm]{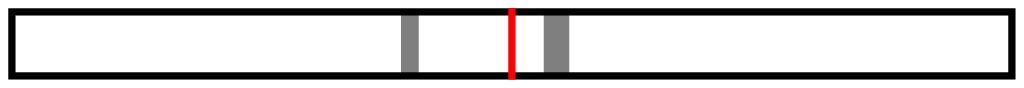} \\
85141	&	HD 157498	&			9.29	 		&		0.58	 		&			1.79	 		&			1.75	 		&		0.87	 	&		0.86	 	&		31.22	 	&	2	&		\includegraphics[width=45mm]{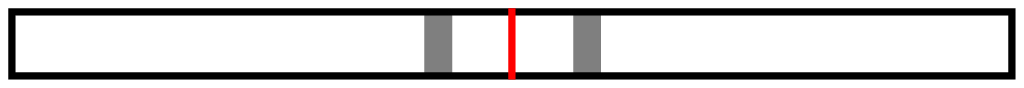} \\
86201	&	HD 160922	&	0.082	&	0	&	1.460	&	1.180	&	0.02	&	0.011	&	0.143	&	12	&		\includegraphics[width=45mm]{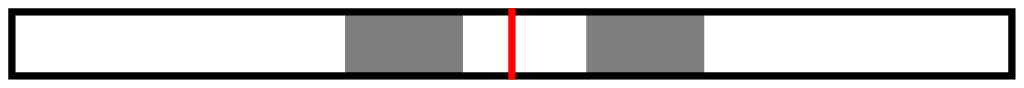} \\
\\
86221	&	WDS 17370+2753	&			9	 		&		0.21	 		&			0.64	 		&			0.63	 		&		1.8	 	&		1.79	 	&		26.4	 	&	6	&		\includegraphics[width=45mm]{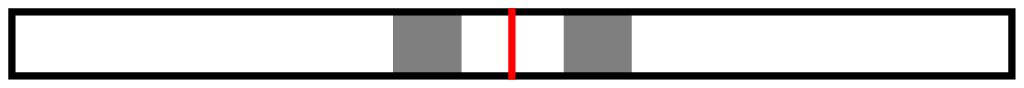} \\
86400	&	HD 1360346	&			0.39	 		&		0.23	 		&			0.72	 		&			0.39	 		&		0.08	 	&		0.06	 	&		1.15	 	&	1	&		\includegraphics[width=45mm]{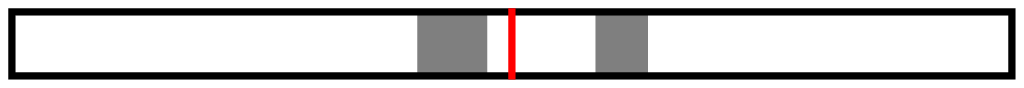} \\
86722	&	HD 161198	&			3.97	 		&		0.94	 		&			0.94	 		&			0.34	 		&		0.04	 	&		0.03	 	&		14.21	 	&	2	&		\includegraphics[width=45mm]{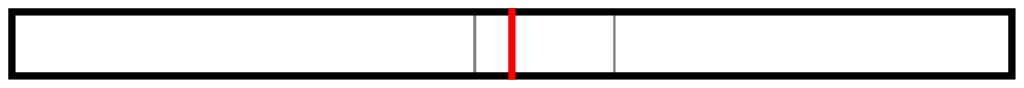} \\
86974	&	HD 161797   	&			22	 		&		0.32	 		&			1.15	 		&			0.13	 		&		4.92	 	&		1.85	 	&		65.17	 	&	4	&		\includegraphics[width=45mm]{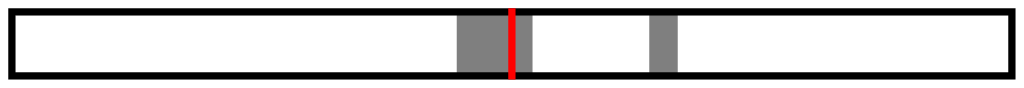} \\
87895	&	HD 163840	&			2.14	 		&		0.41	 		&			0.99	 		&			0.68	 		&		0.32	 	&		0.27	 	&		6.84	 	&	1	&		\includegraphics[width=45mm]{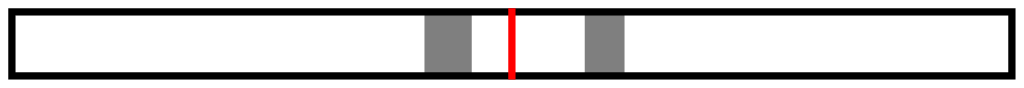} \\

\hline
\end{tabular}
\end{figure*}
\end{landscape}

\begin{landscape}
\begin{figure*}
\begin{tabular}{llllllllllc}
\hline
Object		&Alter. name&	 $a$	&      $e$	&    	$M_{1}$		&  	$M_{2}$		&	$R{ce1}$	&	$R{ce2}$	&	$R_{cb}$	&Ref&	Scheme	\\
 HIP		&	&       (AU)	&		&	(M$_{\odot}$)		&	(M$_{\odot}$)		&	 (AU)		&  	(AU)		&  	(AU)		& &  	 $*_{1}$  \   $*_{2}$ \\
\hline
89937	&	HD 170153	&			1.05	 		&		0.41	 		&			1.18	 		&			0.77	 		&		0.16	 	&		0.13	 	&		3.35	 	&	1	&		\includegraphics[width=45mm]{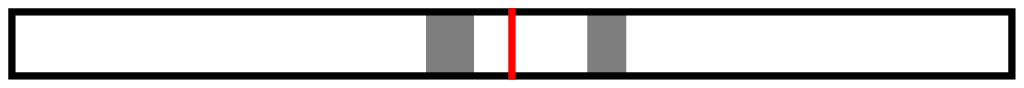} \\
90355	&	HD 169822   	&			0.84	 		&		0.48	 		&			0.91	 		&			0.3	 		&		0.12	 	&		0.07	 	&		2.71	 	&	4	&		\includegraphics[width=45mm]{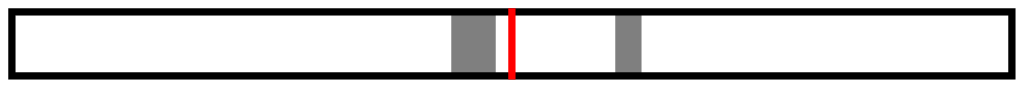} \\
91768	&	HD 173739	&			49.51	 		&		0.53	 		&			0.39	 		&			0.34	 		&		5.43	 	&		5.1	 	&		164.2	 	&	3	&		\includegraphics[width=45mm]{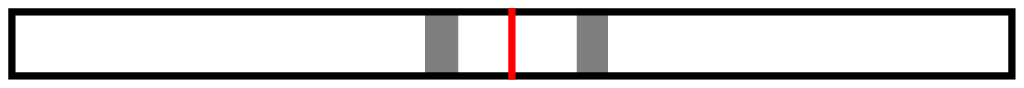} \\
92418	&	HD 174457   	&			1.9	 		&		0.23	 		&			1.07	 		&			0.06	 		&		0.52	 	&		0.14	 	&		5.26	 	&	4	&		\includegraphics[width=45mm]{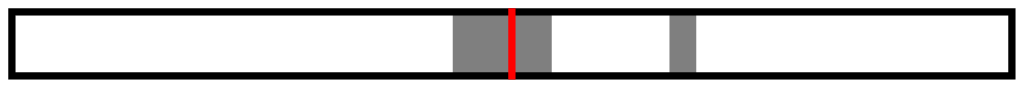} \\
92835	&	HP Dra	&	0.123	&	0.06	&	1.102	&	1.099	&	0.03	&	0.03	&	0.31	&	11	&		\includegraphics[width=45mm]{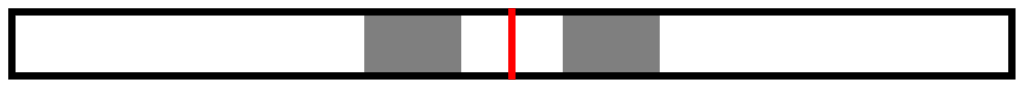} \\
\\
\bf{93017}&	\bf{ADS 11871}	&	\bf{22.96}	&		\bf{0.25} 		&	\bf{1.65}	 	&			\bf{1.58}	 		&		\bf{4.34}	 	&		\bf{4.26}	 	&		\bf{68.77}	 	&	\bf{5}	&		\includegraphics[width=45mm]{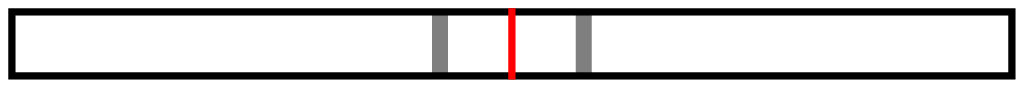} \\
93506	&	WDS 19026+2953 	&			13.36	 		&		0.2	 		&			2.97	 		&			2.4	 		&		2.81	 	&		2.55	 	&		38.93	 	&	6	&		\includegraphics[width=45mm]{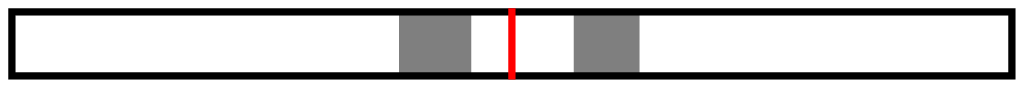} \\
93574	&	HD 175986	&			9.01	 		&		0.39	 		&			1.89	 		&			1.65	 		&		1.35	 	&		1.27	 	&		28.62	 	&	2	&		\includegraphics[width=45mm]{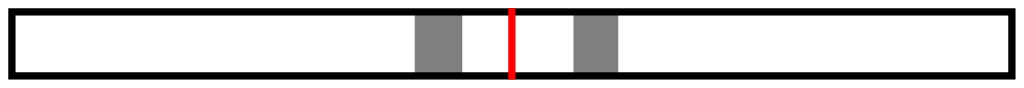} \\
95028	&	HD 181602	&			0.85	 		&		0.37	 		&			1.4	 		&			0.5	 		&		0.15	 	&		0.1	 	&		2.65	 	&	1	&		\includegraphics[width=45mm]{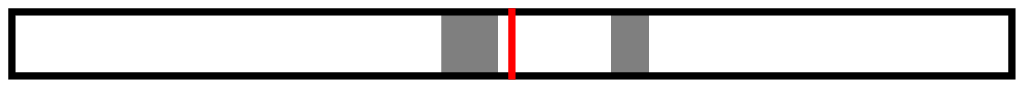} \\
95575	&	HD 183255	&			0.62	 		&		0.15	 		&			0.78	 		&			0.38	 		&		0.15	 	&		0.11	 	&		1.74	 	&	1	&		\includegraphics[width=45mm]{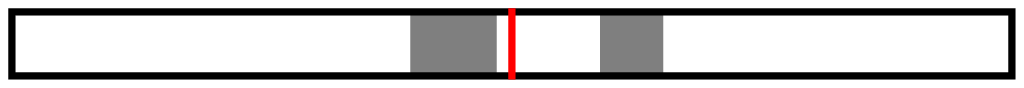} \\
\\
95995	&	HD 184467	&			1.45	 		&		0.37	 		&			1.22	 		&			0.46	 		&		0.26	 	&		0.17	 	&		4.53	 	&	2	&		\includegraphics[width=45mm]{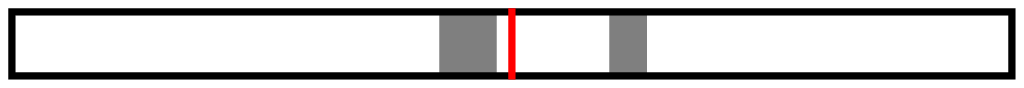} \\
96302	&	HD 184759	&			4.68	 		&		0.82	 		&			3.34	 		&			1.59	 		&		0.18	 	&		0.13	 	&		16.48	 	&	2	&		\includegraphics[width=45mm]{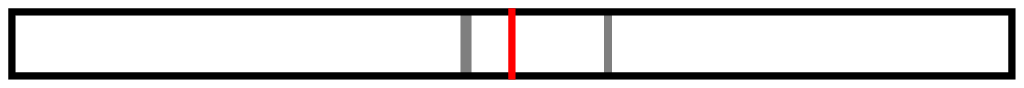} \\
96471	&	HD 184860   	&			1.4	 		&		0.67	 		&			0.77	 		&			0.03	 		&		0.14	 	&		0.03	 	&		4.42	 	&	4	&		\includegraphics[width=45mm]{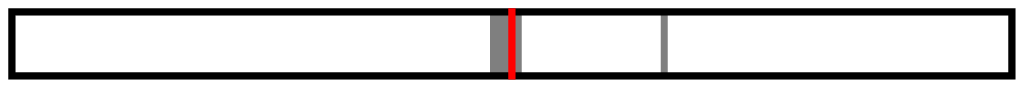} \\
\bf{98001}	&	\bf{HD 188753}	&		\bf{11.65} 		&		\bf{0.47} 		&	\bf{1.3} 	&			\bf{1.11}	 		&		\bf{1.48}	&	\bf{1.38}	 	&		\bf{37.98}	 	&	\bf{2}	&		\includegraphics[width=45mm]{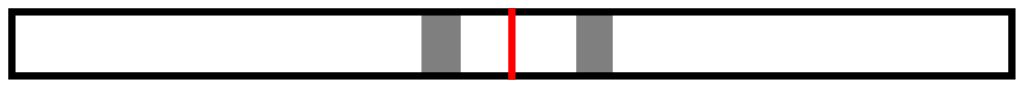} \\
99965	&	HD 193216	&			1.24	 		&		0.08	 		&			0.88	 		&			0.56	 		&		0.32	 	&		0.26	 	&		3.26	 	&	1	&		\includegraphics[width=45mm]{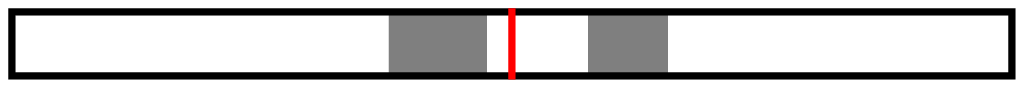} \\
\\
103641	&	HD 200077	&	0.587	&	0.66	&	1.186	&	0.941	&	0.04	&	0.04	&	2.01	&	12	&		\includegraphics[width=45mm]{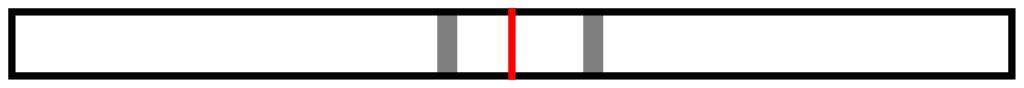} \\
104019	&	WDS 21044+1951	&			12.86	 		&		0.39	 		&			1.67	 		&			1	 		&		2.06	 	&		1.63	 	&		40.74	 	&	6	&		\includegraphics[width=45mm]{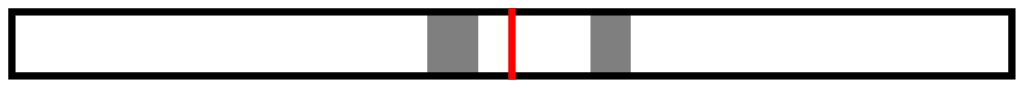} \\
105969	&	HD 204613	&			2.06	 		&		0.13	 		&			1.01	 		&			0.49	 		&		0.52	 	&		0.38	 	&		5.68	 	&	1	&		\includegraphics[width=45mm]{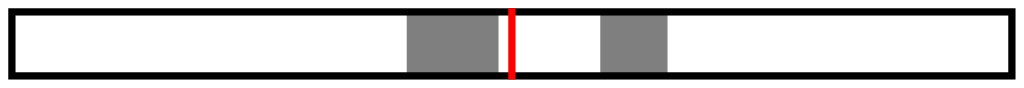} \\
107354	&	HD 206901	&			8.24	 		&		0.31	 		&			1.56	 		&			2.6	 		&		1.53	 	&		1.21	 	&		25.32	 	&	2	&		\includegraphics[width=45mm]{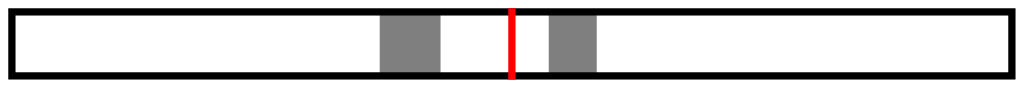} \\
108473	&	HD 208776   	&			4.2	 		&		0.27	 		&			1.14	 		&			0.51	 		&		0.87	 	&		0.61	 	&		12.62	 	&	4	&		\includegraphics[width=45mm]{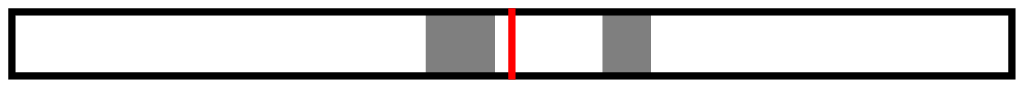} \\
\\
109176	&	HD 210027	&			0.12	 		&		0	 		&			1.25	 		&			0.8	 		&		0.03	 	&		0.03	 	&		0.22	 	&	1	&		\includegraphics[width=45mm]{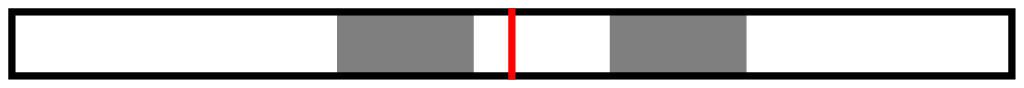} \\
110893	&	HD 239960	&			9.51	 		&		0.42	 		&			0.28	 		&			0.15	 		&		1.46	 	&		1.1	 	&		30.4	 	&	3	&		\includegraphics[width=45mm]{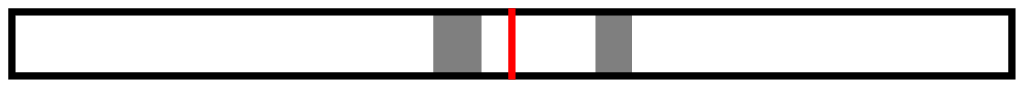} \\
110893	&	ADS15972 		&		9.53	 		&		0.41	 		&			0.27	 		&			0.17	 		&		1.51	 	&		1.20	 	&		30.41	 	&	5	&		\includegraphics[width=45mm]{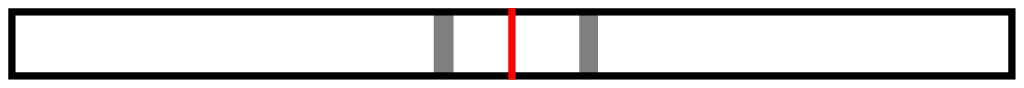} \\
111170	&	HD 213429	&			1.74	 		&		0.38	 		&			1.08	 		&			0.7	 		&		0.28	 	&		0.23	 	&		5.5	 	&	1	&		\includegraphics[width=45mm]{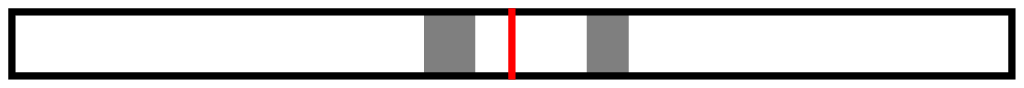} \\
113718	&	HD 217580	&			1.16	 		&		0.54	 		&			0.76	 		&			0.18	 		&		0.15	 	&		0.08	 	&		3.78	 	&	1	&		\includegraphics[width=45mm]{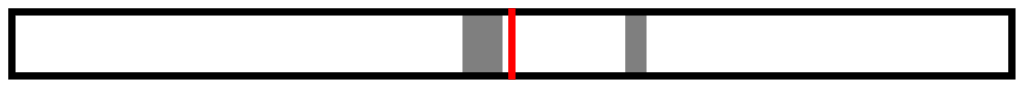} \\

\hline
\end{tabular}
\end{figure*}
\end{landscape}

\subsection{Application to Real Systems}

\label{examples}
The perturbing effect of stellar companions lead to a widespread
belief that the presence of planets in binary systems was very
unlikely. Now we know that binaries can have planets, and thus should
have stable regions around them where planet formation took
place. Queloz \et 2000, and Hatzes \et 2003, discovered two
giant planets in the binary systems GJ 86 and $\gamma$Cephei. Since
then, binaries have become important targets in the search for
extrasolar planets, particularily given their abundance.

Until now about 70 binary systems with planets have been discovered
(Wright \et 2011), but only for eight of them, orbital parameters
(semimajor axis, eccentricity and mass ratio) are available (Desidera
\& Barbieri 2007, Doyle \et 2011, Muterspaugh \et 2010). Table
\ref{tabla_muestra} shows our prediction for the extent of the stable
orbits regions for these systems (in bold type letter). In all cases,
where the semimajor axis of the planet is known, the observed planet
is located within the predicted stable zone. We present figures with
the calculated stable regions constructed with invariant loops, for
the five cases where planets are confirmed, and the values for the
semimajor axis of observed planets are known. In particular, notice
the case of HD 120136, this is an open binary with a very high
eccentricity, usually treated as single star for this reason. The
large eccentricity results in a very narrow, circumstellar stable
region. Even so, the discovered planet lies in a P-type orbit inside
our predicted circumstellar stable region.

\subsubsection{\bf{HIP 10138 (HD 13445 or GL 86)}}
This binary system is located at a distance of 10.9 $\pm$ 0.08 pc.
The companion of this object, discovered by Els \et 2001, has a
semimajor axis of 18.4 AU, with an eccentricity of 0.4. The spectral
type of the most massive component is K1 with a mass of 0.77 M$_\odot$
and it is a white dwarf (Mugrauer \& Neuhauser 2005). The companion
has a mass of 0.49 M$_\odot$. At the moment, one planet was found in
this system with a mass of $M_{p}\sin i$ = 4.0 M{$_{\rm J}$}. It has a
semimajor axis of 0.113 AU and an eccentricity of 0.042 (Bonavita \&
Desidera 2007). For this binary the calculated stable zone located
around each star is $R_{ce1}$ = 3.06 AU around the principal
component, $R_{ce2}$ = 2.49 AU around the companion, in both cases
this radii is the outermost radii possible to have stable orbits, and
$R_{cb}$ = 58.53 AU as the innermost radii available for circumbinary
orbits.

\begin{figure}
\includegraphics[width=74mm]{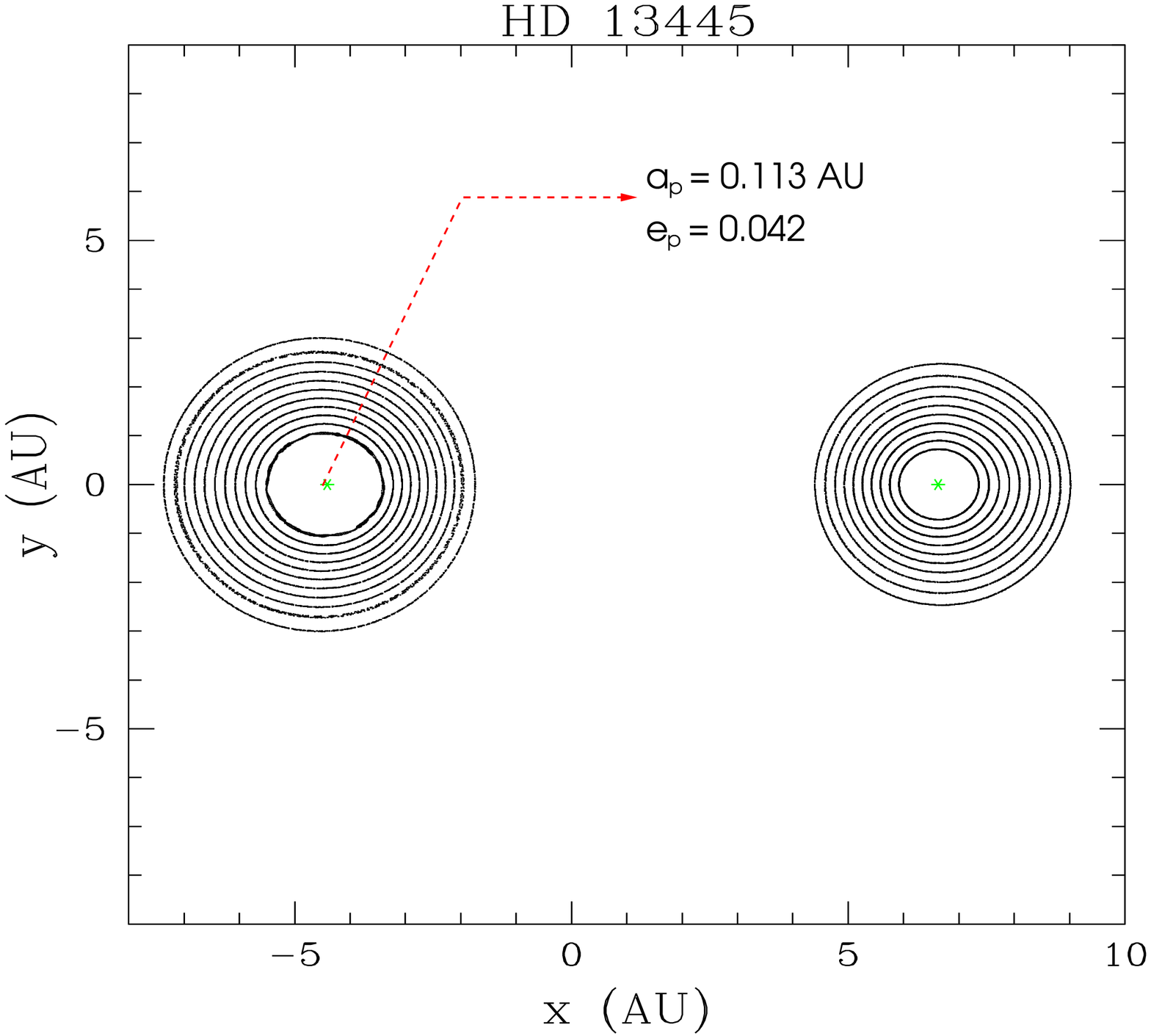}
\includegraphics[width=74mm]{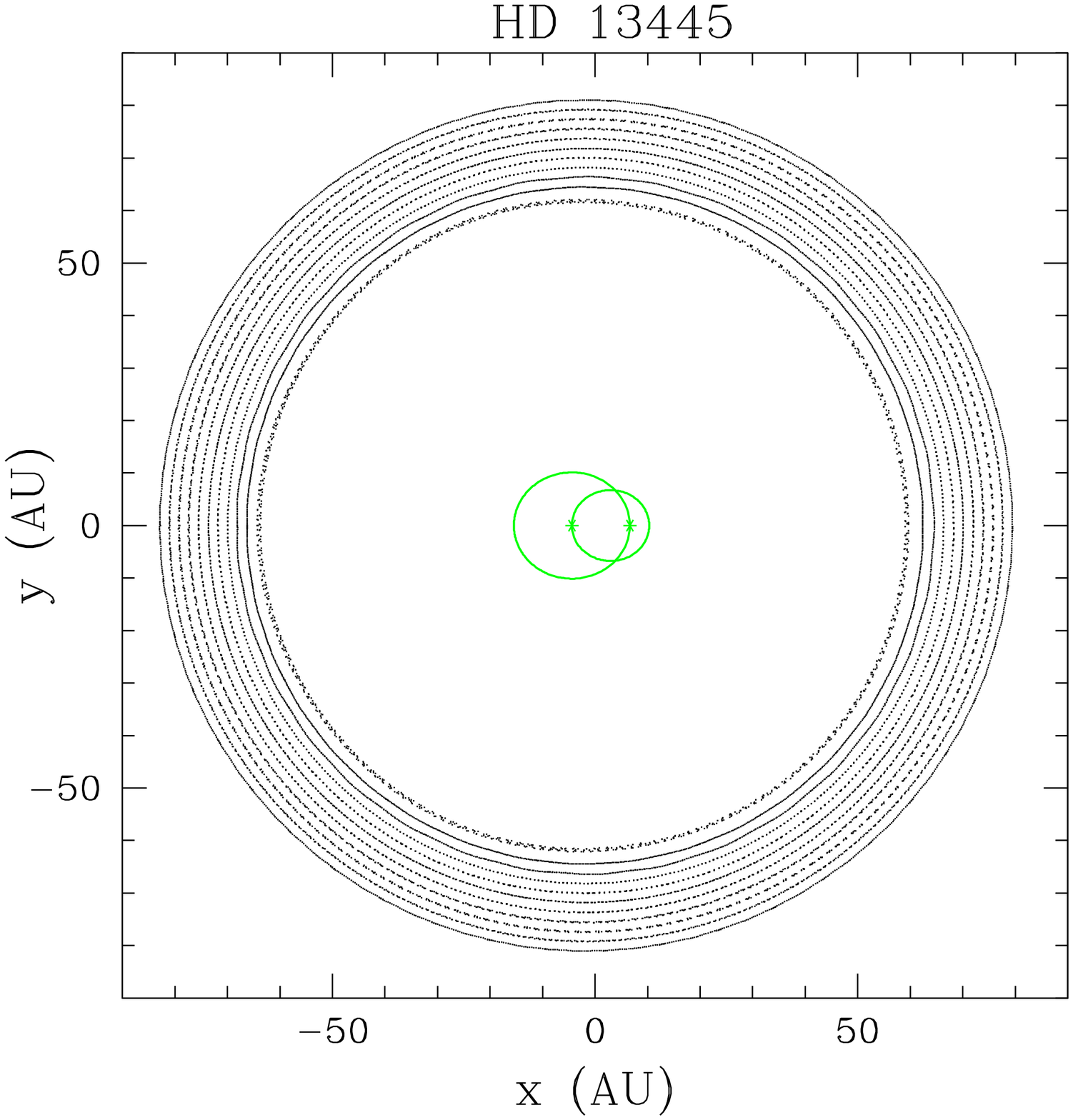}
\vspace{0.1cm}
\caption {Stable zones around the system HD 13445. The upper panel
  shows the circumstellar regions (primary to the left) and the bottom
  panel the circumbinary region. Notice the change in scale. The green
  lines show the stellar orbits, the barycenter is at the center and
  the system is shown at periastron. Orbital parameters: M$_1$ = 0.77
  M$_\odot$, M$_2$ = 0.49 M$_\odot$, $e$ = 0.4, $a$ = 18.4 AU}
\label{fig.HD13445}
\end{figure}

In Figure \ref{fig.HD13445}, we present the circumprimary,
circumsecondary and circumbinary regions of orbital stability, for
planets in this case, to settle down, calculated at periastron with
our method. The stellar orbits are marked in green. The known planet
is orbiting the primary star in a very small orbital radius,
indistinguishable in this figure.

\subsubsection{\bf{HIP 14954 (HD 19994 or 94 Cet)}}
This binary system is located at a distance of 22.6 pc. Its semimajor
axis is 120 AU and it has eccentricity of 0.26. The mass of the
primary star is 1.35 M$_\odot$ with a spectral type F8 V, the mass of
the secondary star is 0.35 M$_\odot$. The planet orbiting this object
has a semimajor axis of 1.428 AU with an eccentricity of 0.30 and
$M_p\sin\ i$= 1.69 M$_{\rm J}$ (Desidera \& Barbieri 2007).

\begin{figure}
\includegraphics[width=74mm]{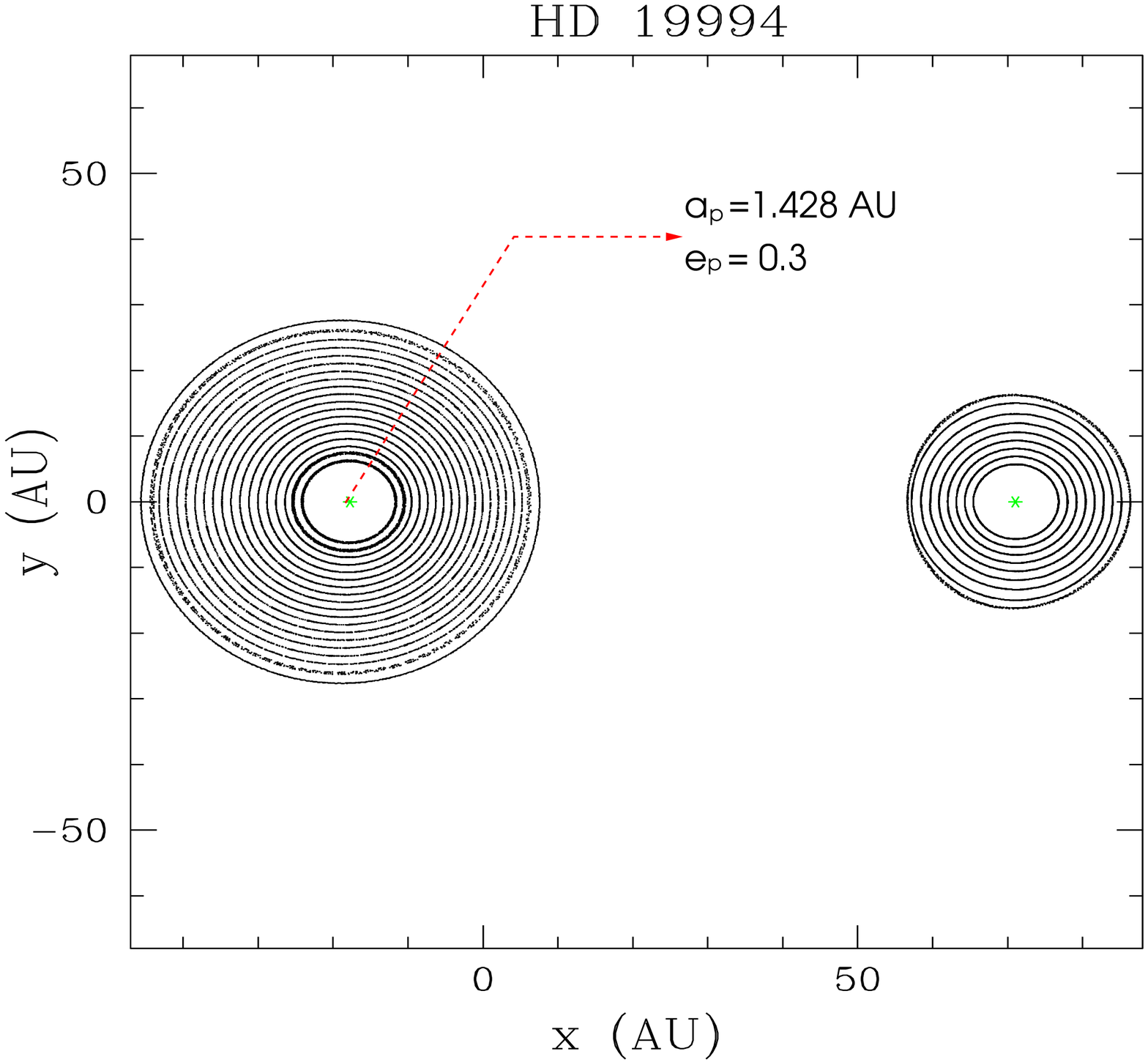}
\includegraphics[width=74mm]{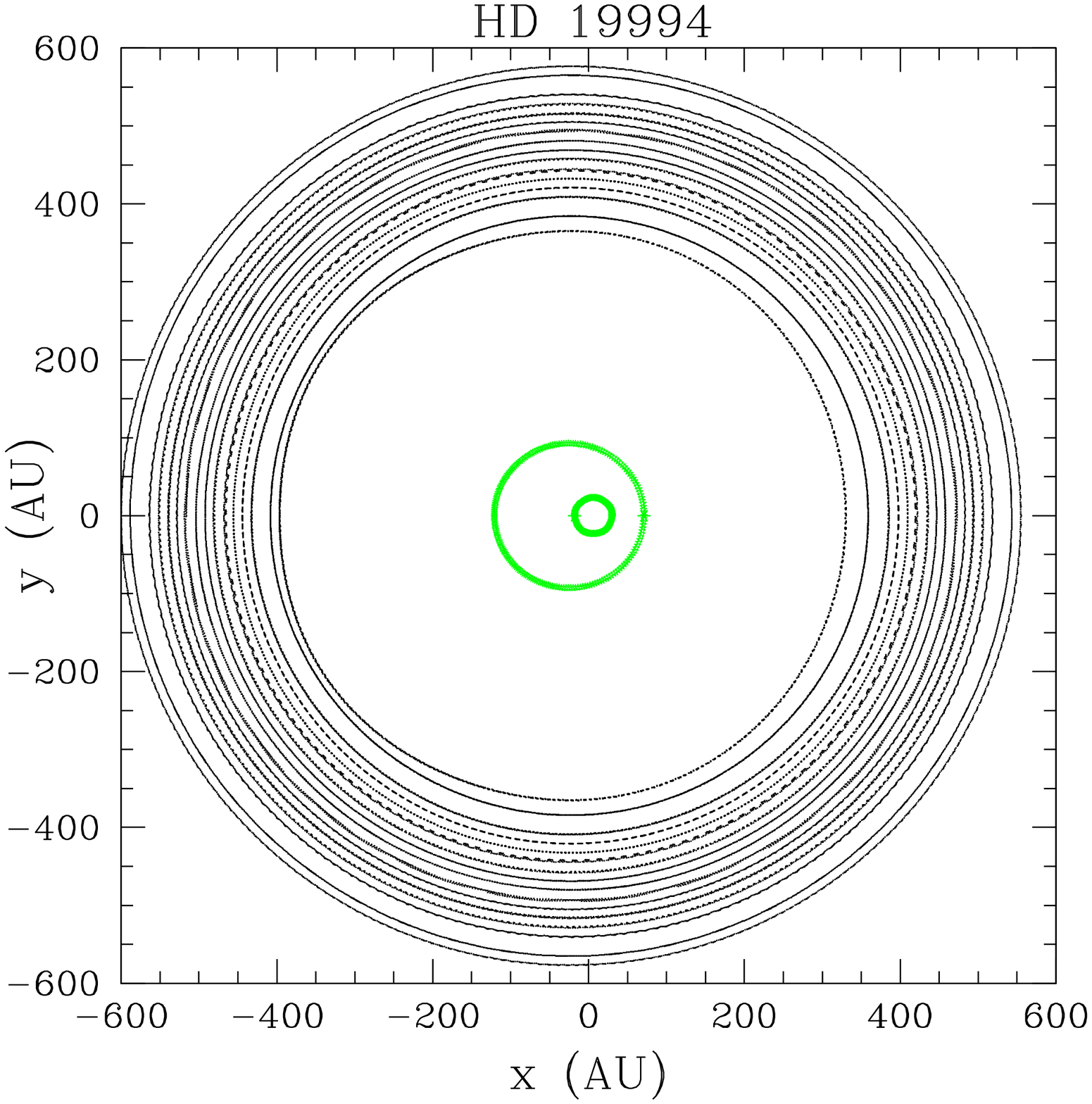}
\vspace{0.1cm}
\caption {Stable zones around the system HD 19994 (94 Cet). The upper
  panel shows the circumstellar regions, primary star to the left
  secondary to the right. The bottom panel shows the circumbinary
  region, notice the change in scale. The green curves show the
  stellar orbits, the system is presented at periastron. Orbital
  parameters: $M_1$ = 1.35 M$_\odot$, $M_2$ = 0.35 M$_\odot$, $e$ = 0.26,
  $a$ = 120 AU}
\label{fig.HD19994}
\end{figure}

In Figure \ref{fig.HD19994}, circumstellar, circumbinary stable
regions, and planetary orbit (in red) are shown. Stellar orbits
(green) are also indicated.

\subsubsection{\bf{HIP 67275 (HD 120136 or $\tau$ Boo)}}

This system is located at 15.62 pc, and it has the largest
eccentricity for the cases of binary systems with known orbital
parameters, which here are semimajor axis 245 AU, eccentricity of 0.91
and masses 1.35 M$_\odot$ for the primary and 0.4 M$_\odot$ for the
secondary component. The planet observed in this system has a
semimajor axis of 0.048 AU while our approach predicts a maximum radii
of 4.86 AU, the eccentricity planet is 0.023 and $M_p\sin i$ is 4.13
(Desidera \& Barbieri 2007).

\begin{figure}
\includegraphics[width=74mm]{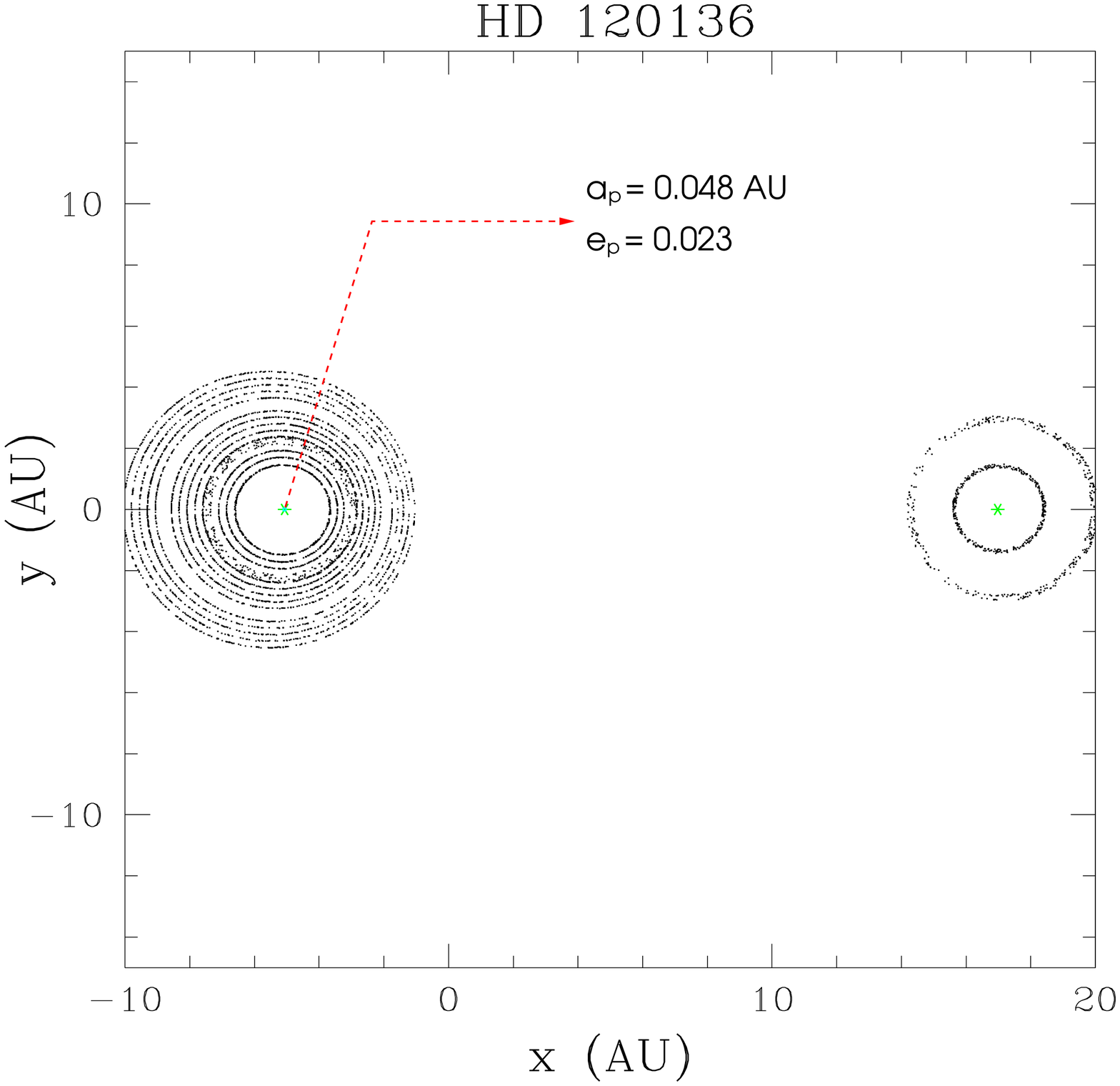}
\includegraphics[width=74mm]{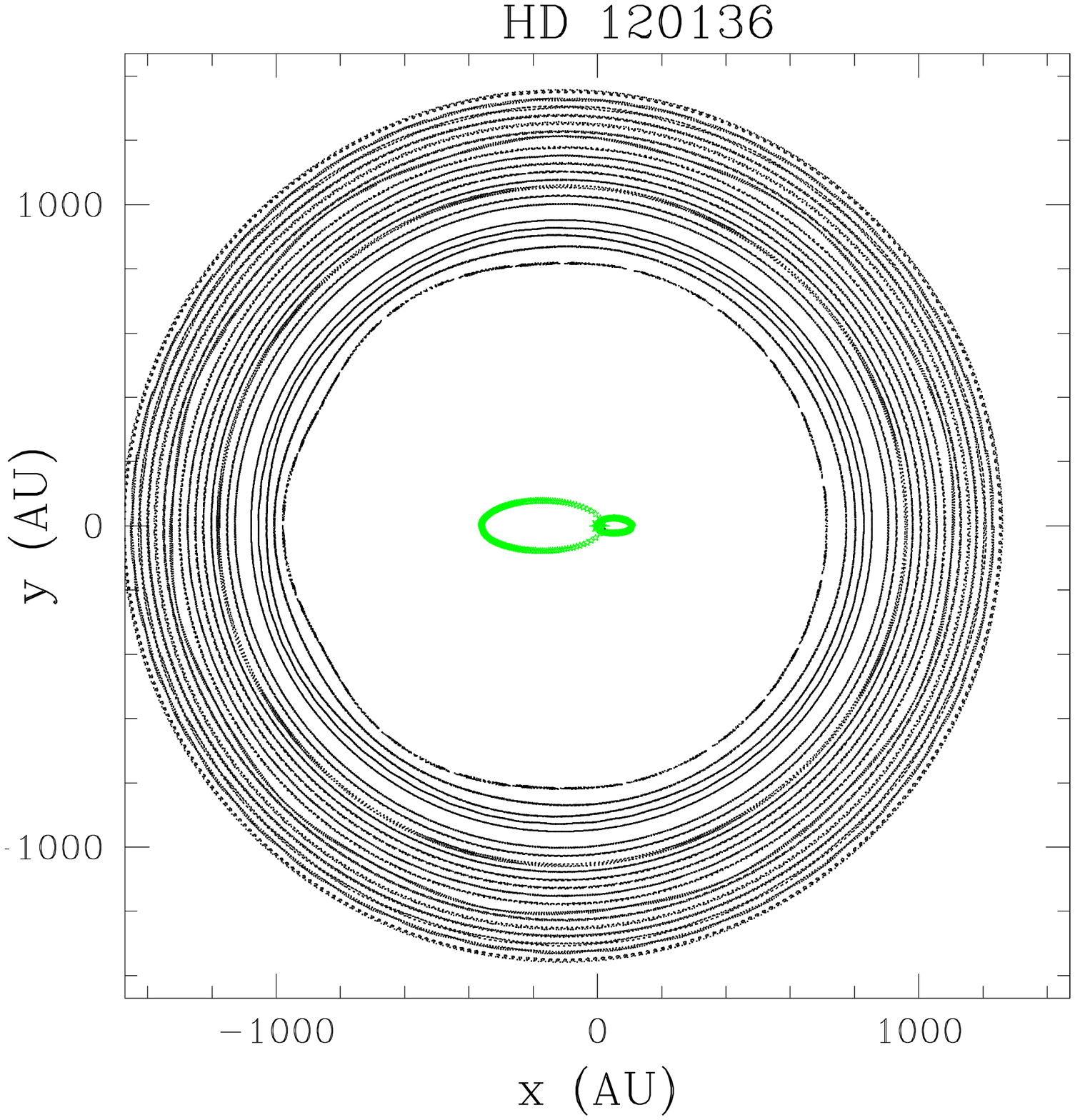}
\vspace{0.1cm}
\caption {Stable zones around the system HD 120136 ($\tau$ Boo). The
  upper panel shows the circumstellar regions, primary star to the left
  secondary to the right. The bottom panel shows the circumbinary
  region, notice the change in scale. The green curves show the
  stellar orbits, the system is presented at periastron. Orbital
  parameters: $M_1$ = 1.35 M$_\odot$, $M_2$ = 0.4 M$_\odot$, $e$ = 0.91,
  $a$ = 245 AU}
\label{fig.HD120136}
\end{figure}

In Figure \ref{fig.HD120136}, we present the circumprimary,
circumsecondary and circumbinary stable regions, for planets in this
case, calculated at periastron. The stellar orbits are marked in
green. The known planet is orbiting the primary star in a very small
orbital radius, indistinguishable in this figure.

\subsubsection{\bf{HIP 116727 (HD 222404 or Gamma Cep)}}
This system is located at a distance of 14.1 pc. The spectral type of
the main component is K, and its mass is 1.59M$_\odot$, while the mass
of the companion is 0.4 M$_\odot$, the semimajor axis is 18.5 AU with
an eccentricity of 0.36. The planet in this system has a semimajor
axis of 2.14 AU, with an eccentricity 0.12 and $M_p\sin i$
1.77 M{$_{\rm J}$} (Desidera \& Barbieri 2007).

\begin{figure}
\includegraphics[width=74mm]{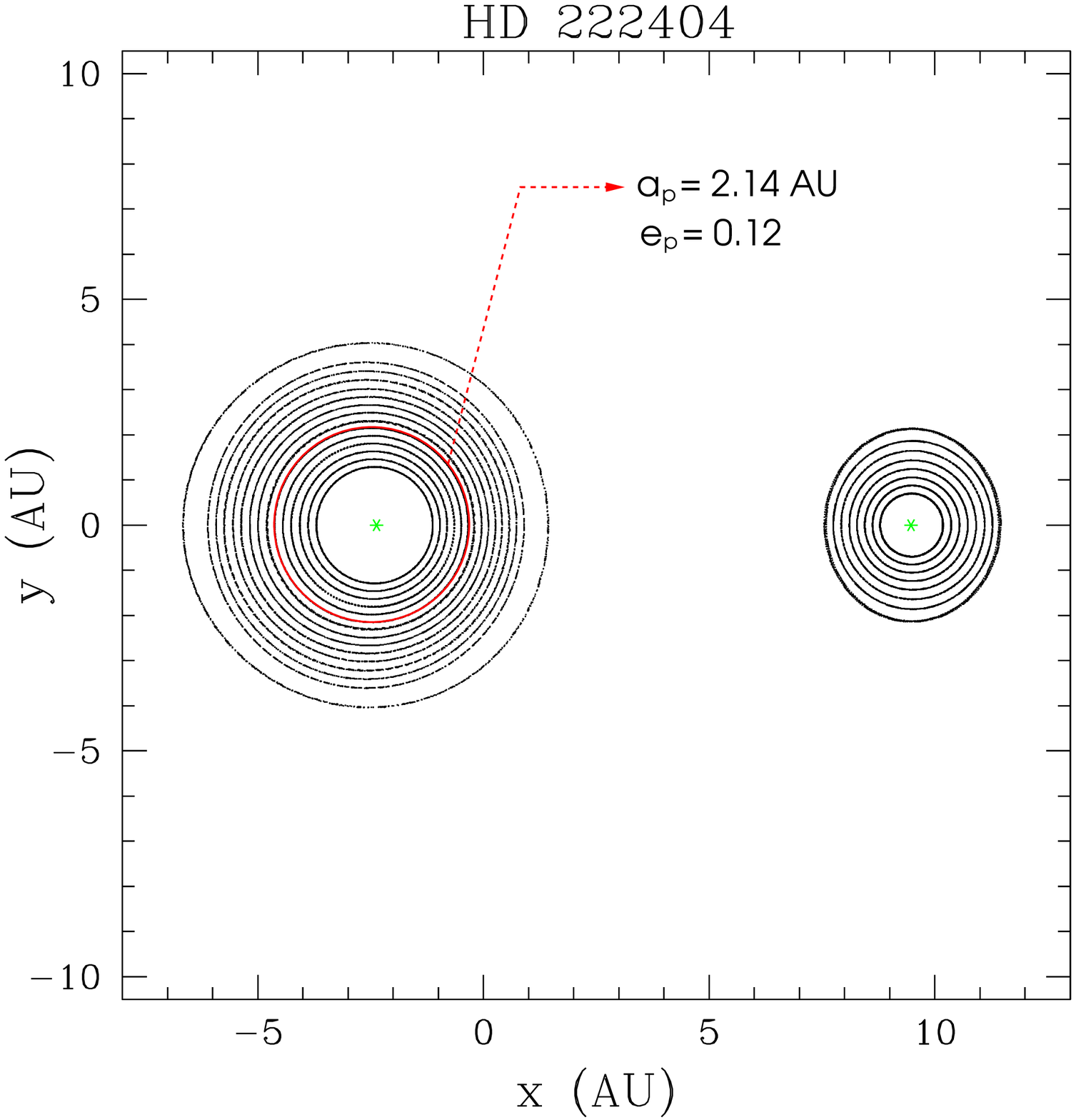}
\includegraphics[width=74mm]{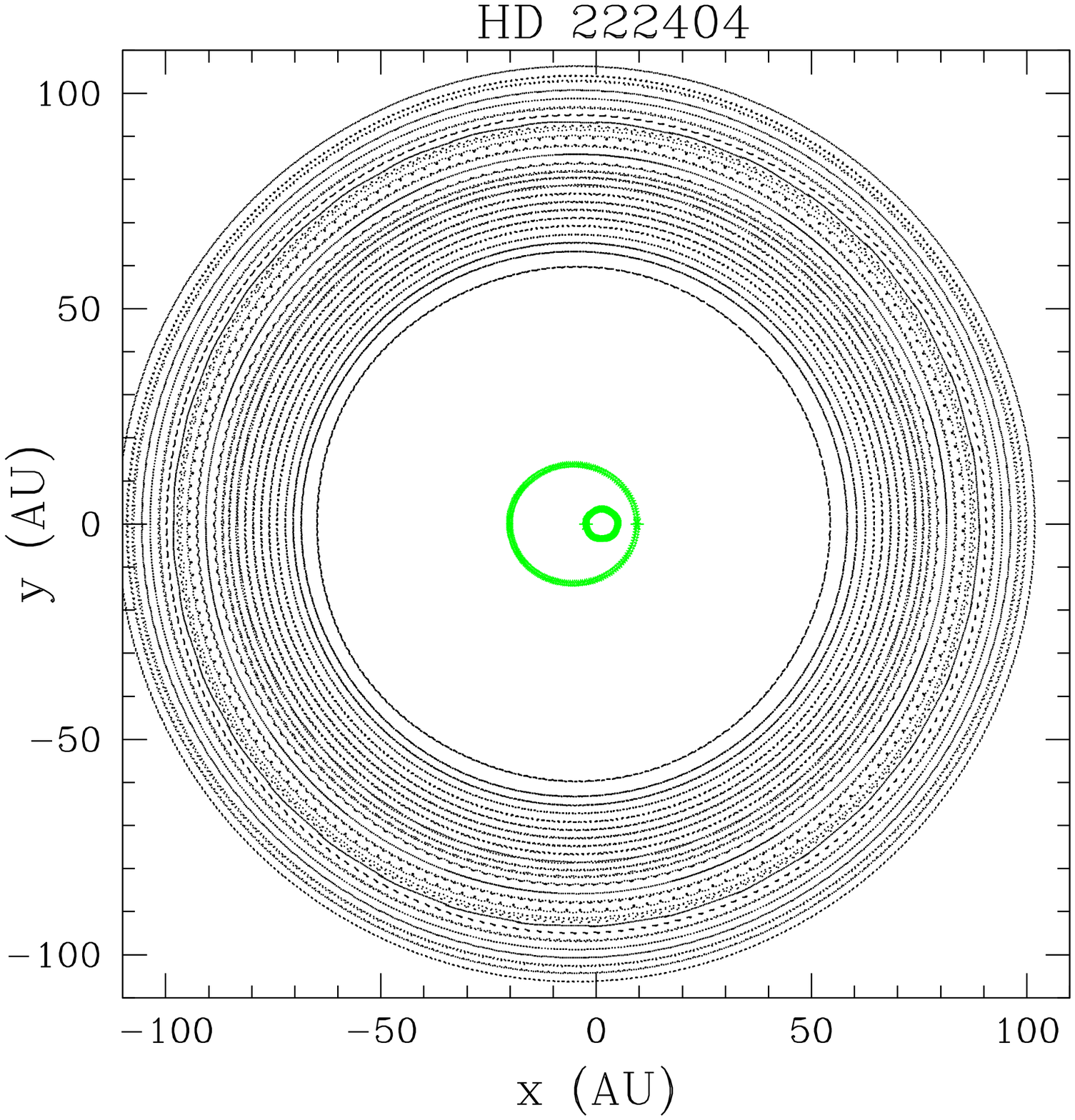}
\vspace{0.1cm}
\caption {Stable zones around the system HD 222404 (Gamma Cep). The
  upper panel shows the circumstellar regions, primary star to the
  left secondary to the right. The bottom panel shows the circumbinary
  region, notice the change in scale. The green curves show the
  stellar orbits, the system is presented at periastron. In this case
  planet orbit is not to close to the star, line darker (red) around
  the primary star shows its orbit. Orbital parameters for the star:
  $M_1$ = 1.59 M$_\odot$, $M_2$ = 0.4 M$_\odot$, $e$ = 0.36, $a$ = 18.5
  AU}.
\label{fig.HD222404}
\end{figure}

In Figure \ref{fig.HD222404}, circumstellar stable regions,
circumbinary stable region (gray), and planetary orbit (in red) are
shown. Stellar orbits (green) are also indicated.

\subsubsection{\bf{Kepler 16}}
Recently it was found in this system a planet in circumbinary orbit,
this the first observed of this kind (on the circumbinary disc). The
primary star has a mass of 0.69 M$_\odot$, and the companion has a
mass of 0.20 M$_\odot$, the semimajor axis is 0.22 AU with an
eccentricity of 0.16 (Doyle \et 2011). The discovered planet has a
semimajor axis of 0.71 AU, with an eccentricity of 0.0069 and a mass
of 0.33 M{$_{\rm J}$}.

\begin{figure}
\includegraphics[width=73mm,height=75mm]{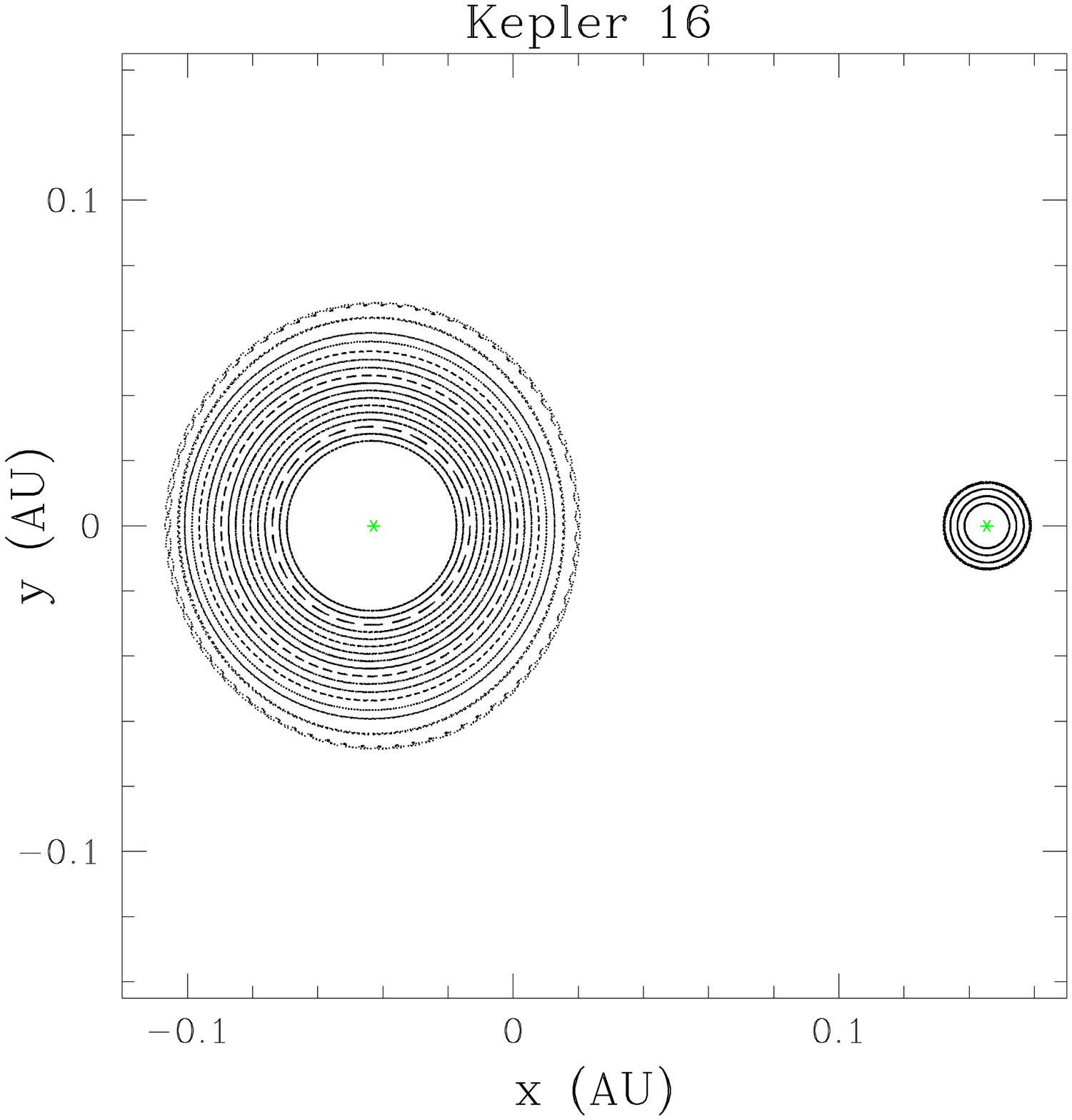}
\includegraphics[width=73mm,height=75mm]{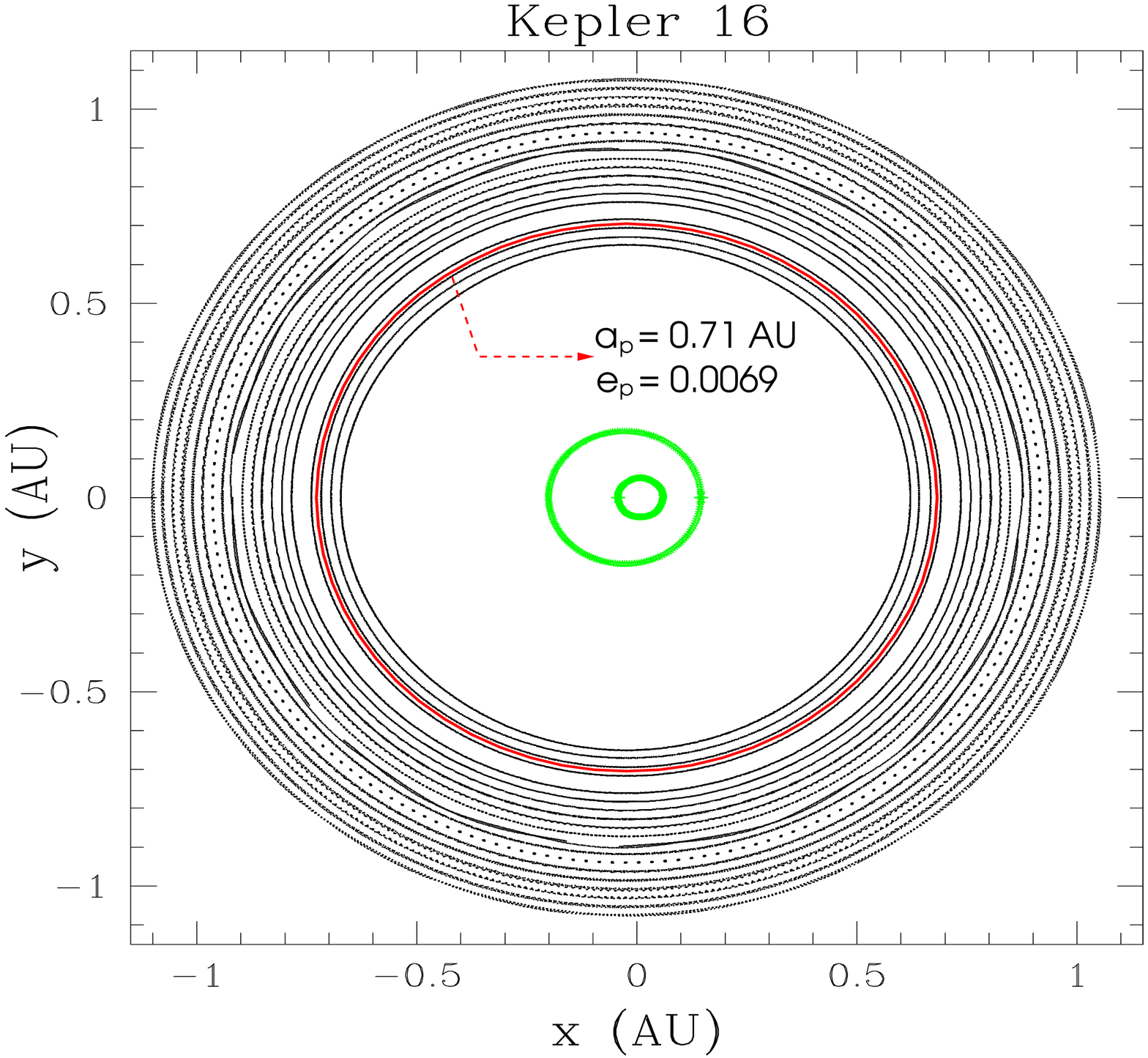}
\vspace{0.1cm}
\caption {Stable zones around the system Kepler 16. The upper panel
  shows the circumstellar regions, primary star to the left secondary
  to the right. The bottom panel shows the circumbinary region, notice
  the change in scale. The green curves show the stellar orbits, the
  system is presented at periastron. Orbital parameters: $M_1$ = 0.69
  M$_\odot$, $M_2$ = 0.20 M$_\odot$, $e$ = 0.16, $a$ = 0.22 AU}
\label{fig.Kepler16}
\end{figure}

In Figure \ref{fig.Kepler16}, circumstellar stable regions,
circumbinary stable region, and planetary orbit (in red) are
shown. Stellar orbits (green) are also indicated.

\section{Conclusions}\label{conclusions}
We have compiled a sample of binary stars with known orbital
parameters (semimajor axes, eccentricities and stellar masses) of the
Solar neighborhood and present some basic statistics.

We calculate on this binary stars sample the extent of regions of
stable non-self intersecting orbits where planets may exist. For this
purpose, we have applied the concept of ``invariant loops'' and used
the formulae of PSA1 and PSA2. Our approximation is ballistic, thus,
the application is straightforward to debris discs (planets, cometary
nucleii, asteroid belts, etc.). In the case of gas discs, further
physics may constraint discs sizes, however, what we are providing
here are the regions where the most important orbits of the binary,
i.e., the ones that represent the backbone of the dynamical system
(the ones that are followed for the most of the orbits), lay. We have
computed the spatial limits of these circumstellar and circumbinary
zones for a sample of 161 binaries in the Solar neighborhood where
orbital data is known and presented it in the form of a table where
all the relevant parameters are provided.

We compare our results with observations in the 5 cases where planets
have been discovered in binary systems, and where semimajor axis for
the planets are provided. We find that all the planets lay down within
our computed regions of stability. In particular, for HD 120136, our
predicted region of circumstellar stability is very small, and yet the
discovered planet lays within it. Although confrontation with a larger
database is desirable, the current statistics is fully consistent with
our results, proving reliable our approach. The tool of ``invariant
loops'' may be very helpful in the search for planets in binary
systems.

\section*{Acknowledgments}
We thank our referee, Rudolf Dvorak for the review of this work
that greatly improved it. We also thank Linda Sparke and Antonio
Peimbert for enlightening discussions and Gustavo Arciniega for
technical support. We thank project UNAM through grant DGAPA-PAPIIT
IN110711-2.

\label{lastpage}


\begin{thebibliography}{}

\bibitem[\protect\citeauthoryear{Alcock \et}{2001}]{AAA01} 
Alcock, C.; Allsman, R. A.; Alves, D. R.; Axelrod, T. S.; Becker,
A. C.; Bennett, D. P.; Cook, K. H.; Drake, A. J.; Freeman, K. C.;
Geha, M.; Griest, K.; Lehner, M. J.; Marshall, S. L.; Minniti, D.;
Nelson, C. A.; Peterson, B. A.; Popowski, P.; Pratt, M. R.; Quinn,
P. J.; Stubbs, C. W.; Sutherland, W.; Tomaney, A. B.; Vandehei, T.;
Welch, D. 2001, ApJ, 552, 259

\bibitem[\protect\citeauthoryear{Artymowicz \& Lubow}{1994}]{AL94}
Artymowicz, P. \& Lubow, S. H. 1994, 421, 651

\bibitem[\protect\citeauthoryear{Balega \et}{2006}]{BB06}
Balega, I. I.; Balega, Y. Y.; Hofmann, K.-H.; Malogolovets, E. V.;
Schertl, D.; Shkhagosheva, Z. U.; Weigelt, G.  2006, A\&A, Vol 448, pp. 703

\bibitem[\protect\citeauthoryear{Bate}{1997}]{B97}
Bate, M. R., 1997, MNRAS, 285, 16

\bibitem[\protect\citeauthoryear{Bate \& Bonnell}{1997}]{BB97}
Bate, M. R., \& Bonnell, I. A. 1997, MNRAS, 285, 33

\bibitem[\protect\citeauthoryear{Bonavita \& Desidera}{2007}]{BD07}
Bonavita, M. \& Desidera, S., A\&A, 2007, 468, 721-729

\bibitem[\protect\citeauthoryear{Bonnell \& Bastien}{1992}]{BB92}
Bonnell, I., \& Bastien, P. 1992, IAU Colloquium 135, 32, 206

\bibitem[\protect\citeauthoryear{Borucki}{2010}]{B10} 
Borucki, W.J.  2010, arXiv:1006.2799v1

\bibitem[\protect\citeauthoryear{Cakirli \et}{2009}]{CI09}
Cakirli, O.; Ibanoglu, C.; Bilir, S.; Sipahi, E. 2009, MNRAS, Vol. 395,
pp. 1649

\bibitem[\protect\citeauthoryear{Carpintero \& Aguilar}{1998}]{CA98}
Carpintero, D., Aguilar, L., 1998, MNRAS, 298, 1

\bibitem[\protect\citeauthoryear{Cieza \et}{2009}]{CPA09}
Cieza, Lucas A.; Padgett, Deborah L.; Allen, Lori E.; McCabe, Caer E.;
Brooke, Timothy Y.; Carey, Sean J.; Chapman, Nicholas L.; Fukagawa,
Misato; Huard, Tracy L.; Noriega-Crespo, Alberto; Peterson, Dawn E.;
Rebull, Luisa M. 2009, ApJ, 696L, 84

\bibitem[\protect\citeauthoryear{Correia \et}{2008}]{CUM08}
Correia, A. C. M.; Udry, S.; Mayor, M.; Eggenberger, A.; Naef, D.;
Beuzit, J.-L.; Perrier, C.; Queloz, D.; Sivan, J.-P.; Pepe, F.;
Santos, N. C.; Ségransan, D. 2008, A \& A, 479, 271

\bibitem[\protect\citeauthoryear{Deeg \et}{2008}]{}
Deeg, H. J.; Oca\~na, B.; Kozhevnikov, V. P.; Charbonneau, D.;
O'Donovan, F. T.; Doyle, L. R. 2008, A \& A, 480, 563

\bibitem[\protect\citeauthoryear{Desidera \& Barbieri}{2007}]{DB07}
Desidera, S. \& Barbieri M. 2007, A\&A, 462, 345

\bibitem[\protect\citeauthoryear{D\'\i az \et}{2007}]{DG07}
D\'\i az, R. F.; Gonz\'alez, J. F.; Cincunegui, C.; Mauas, P. J. D.
2007, A\&A, Vol. 474, pp. 345

\bibitem[\protect\citeauthoryear{Di Stefano}{2001}]{DS01}
Di Stefano, R. 2001. See Zinnecker \& Mathieu 2001, pp. 529-38

\bibitem[\protect\citeauthoryear{Dong-Wook \et}{2008}]{DC08} 
Dong-Wook, Lee; Chung-Uk, Lee; Byeong-Gon; Park; Sun-Ju, Chung; 
Young-Soo, Kim; Ho-Il, Kim; Cheongho, Han. 2008, ApJ, 672, 623

\bibitem[\protect\citeauthoryear{Doyle \et}{2011}]{DC11}
Doyle, Laurance R.; Carter, Joshua A.; Fabrycky, Daniel C.; Slawson,
Robert W.; Howell, Steve B.; Winn, Joshua N.; Orosz, Jerome A.; Prˇsa,
Andrej; Welsh, William F.; Quinn, Samuel N.; Latham, David; Torres,
Guillermo; Buchhave, Lars A.; Marcy, Geoffrey W.; Fortney, Jonathan
J.; Shporer, Avi; Ford, Eric B.; Lissauer, Jack J.; Ragozzine, Darin;
Rucker, Michael; Batalha, Natalie; Jenkins, Jon M.; Borucki, William
J.; Koch, David; Middour, Christopher K.; Hall, Jennifer R.;
McCauliff, Sean; Fanelli, Michael N.; Quintana, Elisa V.; Holman,
Matthew J.; Caldwell, Douglas A.; Still, Martin; Stefanik, Robert P.;
Brown, Warren R.; Esquerdo, Gilbert A.; Tang, Sumin; Furesz, Gabor;
Geary, John C.; Berlind, Perry; Calkins, Michael L.; Short, Donald R.;
Steffen, Jason H.; Sasselov, Dimitar; Dunham, Edward W.; Cochran,
William D.; Boss, Alan; Haas, Michael R.; Buzasi, Derek; Fischer,
Debra 2011, Sci. Vol. 333, pp. 1602

\bibitem[\protect\citeauthoryear{Duquennoy \& Mayor}{1991}]{DM91} 
Duquennoy, A. \& Mayor, M. 1991, A\&A, 248, 485

\bibitem[\protect\citeauthoryear{Dvorak, R.}{1986}]{D86} 
Dvorak, R. 1986, A\&A, 167, 379

\bibitem[\protect\citeauthoryear{Eggenberger \et}{2004}]{E04}
Eggenberger, A., Udry, S. \& Mayor, M. 2004, A\&A, 417, 353

\bibitem[\protect\citeauthoryear{Eggleton}{1983}]{E83} 
Eggleton, P. P. 1983, ApJ, 268, 368

\bibitem[\protect\citeauthoryear{Faigler \et}{2011}]{FM11}
Faigler, S. and Mazeh, T. and Quinn, S.N. and Latham, D.W. and Tal-Or,
L.  2011, arXiv astro-ph: 1110.2133.

\bibitem[\protect\citeauthoryear{Fischer}{2008}]{F08}
Fischer, Debra A.; Vogt, Steven S.; Marcy, Geoffrey W.; Butler,
R. Paul; Sato, Bun'ei; Henry, Gregory W.; Robinson, Sarah; Laughlin,
Gregory; Ida, Shigeru; Toyota, Eri; (and 5 coauthors),
2007, ApJ, 669, 1336

\bibitem[\protect\citeauthoryear{Goldstein}{2002}]{G02}
Goldstein, H. 2002, {\it Classical Mechanics}, (3d ed.; Addison Wesley)

\bibitem[\protect\citeauthoryear{Guilloteau}{2001}]{G01}
Guilloteau, S. 2001. See Zinnecker \& Mathieu 2001, pp. 547-54

\bibitem[\protect\citeauthoryear{H\'enon}{1970}]{H70}
H\'enon, M., 1970, A\&A, 9, 24

\bibitem[\protect\citeauthoryear{Holman \& Wiegert}{1999}]{HW99}
Holman M.J. and Wiegert P.A. 1999, AJ, Vol. 117, pp 621

\bibitem[Haghighipour \et (2010)]{HD10} Haghighipour,
  N., Dvorak, R., \& Pilat-Lohinger, E.\ 2010, Astrophysics and Space
  Science Library, 366, 285

\bibitem[Haghighipour et al.(2007)]{HS07}
Haghighipour, N., Sigurdsson, S., Lissauer, J., \& Raymond, S.\ 2007,
arXiv:0704.0832

\bibitem[\protect\citeauthoryear{Jancart \et}{2005}]{JJ05}
Jancart S., Jorissen A., Babusiaux C. and Pourbaix D. 2005, A\&A
vol. 442, pp. 365

\bibitem[\protect\citeauthoryear{Konacki}{2005}]{K05}
Konacki, M. 2005, Nature, 436, 230

\bibitem[Konacki et al.(2010)]{2010ApJ...719.1293K} Konacki, M., 
Muterspaugh, M.~W., Kulkarni, S.~R., 
\& He{\l}miniak, K.~G.\ 2010, ApJ, 719, 1293 

\bibitem[\protect\citeauthoryear{Latham \et}{2002}]{LS02}
Latham, David W.; Stefanik, Robert P.; Torres, Guillermo; Davis,
Robert J.; Mazeh, Tsevi; Carney, Bruce W.; Laird, John B.; Morse, Jon
A.  2002 AJ Vol 124 pp 1144

\bibitem[\protect\citeauthoryear{Launhardt}{2001}]{L01}
Launhardt, R. 2001. See Zinnecker \& Mathieu 2001,
pp. 117-21 

\bibitem[\protect\citeauthoryear{Launhardt \et}{2000}]{LS00}
Launhardt, R., Sargent, A.I., Henning, T., Zylka, R., Zinnecker,
H. 2000. In Birth and Evolution of Binary Stars, IAU Symp. No. 200,
ed. B. Reipurth, H. Zinnecker, pp. 103-5. San Francisco:
Astron. Soc. Pac.

\bibitem[\protect\citeauthoryear{Lubow \& Shu}{1975}]{LS75}
Lubow, S. H. \& Shu, F. H. 1975, ApJ, 198, 383

\bibitem[\protect\citeauthoryear{Lyne \et}{1988}]{LB88} 
Lyne A.G., Biggs J.D., Brinklow A., McKenna J. and Ashworth M., 1988,
Nature 332, 45

\bibitem[\protect\citeauthoryear{Maciejewski \& Sparke}{1997}]{MS97} 
Maciejewski, W., Sparke L. S. 1997, ApJL, 484, 117

\bibitem[\protect\citeauthoryear{Maciejewski \& Sparke}{2000}]{MS00} 
Maciejewski, W., Sparke L. S. 2000, MNRAS, 313, 745

\bibitem[\protect\citeauthoryear{Marcy \& Butler}{1998}]{MB98}
Marcy GW, Butler R.P. 1998. Annu. Rev. Astron. Astrophys. 36:57-98 

\bibitem[\protect\citeauthoryear{Marcy \& Butler}{2000}]{MB00}
Marcy GW, Butler R.P. 2000. Publ. Astron. Soc. Pac. 112:137-40 

\bibitem[\protect\citeauthoryear{Martin \et}{1998}]{Martin98}
Martin, C.; Mignard, F.; Hartkopf, W. I. and McAlister, H. A., A\&A, 1998,
133, 149-162.

\bibitem[\protect\citeauthoryear{Marzari \& Scholl}{2004}]{MS04} 
Marzari, F., \& Scholl, H. 2000, ApJ, 543, 328

\bibitem[\protect\citeauthoryear{Mason \et}{1999}]{Mason99}
Mason, B. D.; Douglas, G. G. and Hartkopf, W. I. ApJ, 1999, 117,1023-1036.

\bibitem[\protect\citeauthoryear{Mathieu}{1994}]{M94}
Mathieu, R. D. 1994, ARA\&A, 32, 465

\bibitem[\protect\citeauthoryear{Mathieu}{2000}]{M00}
Mathieu, R. D.,  Ghez, A. M., Jensen, E. L. N. \& Simon, M. 2000, 
in Protostar and Planets IV, ed. V. Mannings, A. P. Boss \& 
S. S. Russell (Tucson: Univ. Arizona Press) 731

\bibitem[\protect\citeauthoryear{Mathieu \et}{1990}]{ML90}
Robert D. Mathieu; David W. Latham \& R. F. Griffin, 1990, AJ, 100,
1859

\bibitem[\protect\citeauthoryear{Mathieu, Walter \& Myers}{1989}]{MWM89}
Mathieu, R. D., Walter, F. M. \& Myers, P. C. 1989, AJ, 98, 987

\bibitem[\protect\citeauthoryear{Mayor \& Queloz}{1995}]{MQ95}
Mayor M, Queloz D. 1995. Nature 378:355-59
 	
\bibitem[\protect\citeauthoryear{Milone \et}{2005}]{MM05}
Milone, E. F.; Munari, U.; Marrese, P. M.; Williams, M. D.; Zwitter,
T.; Kallrath, J.; Tomov, T. 2005, A\&A, Vol. 441, pp. 605

\bibitem[\protect\citeauthoryear{Mugrauer \et}{2005}]{MN05}
Mugrauer, M.; Neuh\"auser, R.; Seifahrt, A.; Mazeh, T.; Guenther,
E. 2005, A\&A, 440, 1051

\bibitem[Muterspaugh et al.(2010)]{MK10} Muterspaugh, M.~W., Konacki,
  M., Lane, B.~F., \& Pfahl, E.\ 2010, Astrophysics and Space Science
  Library, 366, 77

\bibitem[\protect\citeauthoryear{Muterspaugh \et}{2010}]{ML10}
Muterspaugh, M.~W., Lane, B.~F., Kulkarni, S.~R., et al.\ 2010, AJ, 140, 1657 

\bibitem[\protect\citeauthoryear{Nagel \& Pichardo}{2008}]{NP08}
Nagel, Erick; Pichardo, Barbara, 2008, MNRAS, 384, 548

\bibitem[\protect\citeauthoryear{Paczy\'nski}{1977}]{P77}
Paczy\'nski, B. 1977, ApJ, 216, 822

\bibitem[\protect\citeauthoryear{Padgett \et}{1999}]{PS99}
Padgett, D.L., Brandner, W., Stapelfeldt, R., Strom, S.E., Terebey,
S., Koerner, D. 1999. A.J. 117:1490-1504

\bibitem[\protect\citeauthoryear{Padgett \et}{1997}]{PS97}
Padgett, D.L., Strom, S.E., Ghez, A. 1997. ApJ, 477:705-10

\bibitem[\protect\citeauthoryear{Papaloizou \& Pringle}{1977}]{PP77}
Papaloizou, J., \& Pringle, J. E. 1977, 181, 441 

\bibitem[\protect\citeauthoryear{Pfahl \& Muterspaugh}{2006}]{PM06}
Pfahl,E. \& Muterspaugh M. 2006, ApJ, 652, 1694

\bibitem[\protect\citeauthoryear{Pichardo \et}{2005}]{PSA1}
Pichardo, Barbara; Sparke, Linda S.; Aguilar, Luis A. 2005, MNRAS,
359, 521

\bibitem[\protect\citeauthoryear{Pichardo \et}{2009}]{PSA2}
Pichardo, Barbara; Sparke, Linda S.; Aguilar, Luis A. 2008,
MNRAS, 391, 815

\bibitem[Prato \& Weinberger(2010)]{PW10} Prato, L., \&
  Weinberger, A.~J.\ 2010, Astrophysics and Space Science Library,
  366, 1

\bibitem[\protect\citeauthoryear{Quintana \et}{2007}]{Q07} 
Quintana, Elisa V.; Adams, Fred C.; Lissauer, Jack J.; Chambers, John
E. 2007, ApJ, 660, 807

\bibitem[Quintana \& Lissauer(2010)]{QS10} Quintana, E.~V., \&
  Lissauer, J.~J.\ 2010, Astrophysics and Space Science Library, 366,
  265

\bibitem[\protect\citeauthoryear{Quirrenbach}{2001a}]{Q01a}
Quirrenbach, A. 2001a. See Zinnecker \& Mathieu 2001, pp. 539-46

\bibitem[\protect\citeauthoryear{Quirrenbach}{2001b}]{Q01b}
Quirrenbach, A. 2001b. Annu. Rev. Astron. Astrophys. 39:353-401

\bibitem[\protect\citeauthoryear{Quist \& Lindegren}{2000}]{QL00}
Quist, C.F., Lindegren, L. 2000. Astron. Astrophys. 361:770-80
 
\bibitem[\protect\citeauthoryear{Quist \& Lindegren}{2001}]{QL01}
Quist, C.F., Lindegren, L. 2001. See Zinnecker \& Mathieu 2001, pp. 64-68
 	
\bibitem[\protect\citeauthoryear{Reid \et}{2001}]{R01}
Reid, I.N., Gizis, J.E., Kirkpatrick, J.D., Koerner,
D.W. 2001. A.J. 121:489-502

\bibitem[\protect\citeauthoryear{Raghavan \et}{2006}]{RH06}
Raghavan D., Henry T.J., Mason B.D., Subasavage J.P., Jao W.-C., 
Beaulieu  T.D. and Hambly N.C., 2006, ApJ 646, 523

\bibitem[\protect\citeauthoryear{Rudak \& Paczynski}{1981}]{RP81}
Rudak, B., Paczynski, B. 1981, Acta Astron, 31, 13

\bibitem[\protect\citeauthoryear{Rattenbury 2009}{2009}]{R09}
Rattenbury N.J. 2009, MNRAS, 392, 439

\bibitem[\protect\citeauthoryear{Sigurdsson \& Phinney}{1993}]{SP93}
Sigurdsson S. and Phinney E.S.,  ApJ. 415, 631

\bibitem[\protect\citeauthoryear{Sigurdsson \et}{2003}]{SR03}
Sigurdsson S., Richer H.B., Hansen B.M., Stairs I.H. and
Thorsett S.E., 2003, Science 301, 193

\bibitem[\protect\citeauthoryear{Simon \et}{1999}]{SC99}
Simon, M., Close, L.M., Beck, T.L. 1999. A.J. 117:1375-86

\bibitem[\protect\citeauthoryear{Smith \et}{2000}]{SB00}
Smith, K.W., Bonnell, I.A., Emerson, J.P., Jenness, T. 2000. MNRAS
319:991-1000

\bibitem[\protect\citeauthoryear{S\"oderhjelm}{1999}]{S99}
S\"oderhjelm, S. 1999, A\&A. 341:121-40

\bibitem[\protect\citeauthoryear{Strigachev \& Lampens}{2004}]{SL04}
Strigachev, A. \& Lampens, P. A\&A, 2004, 422, 1023-1029.

\bibitem[\protect\citeauthoryear{Th\'ebault \et}{2004}]{T04} 
Th\'ebault, P., Marzari, F., \& Scholl, H. 2004, A\&A, 427, 1097

\bibitem[\protect\citeauthoryear{Udry \et}{2002}]{UM02}
Udry S., Mayor M., Naef D., Pepe F., Queloz D., Santos N.C. and
Burnet M., A\&A, 390, 267

\bibitem[\protect\citeauthoryear{Wolszczan \& Frail}{1992}]{1992Natur.355..145W} 
Wolszczan, A., \& Frail, D.~A.\ 1992, Nature, 355, 145

\bibitem[\protect\citeauthoryear{Wright \et}{2011}]{W11} 
Wright, J. T.; Fakhouri, O.; Marcy, G. W.; Han, E.; Feng, Y.; Johnson,
John Asher; Howard, A. W.; Fischer, D. A.; Valenti, J. A.; Anderson,
J.; Piskunov, N. 2011, PASP, 123, 412




\end{thebibliography}
\end{document}